\documentclass[aps,prd,twocolumn,amssymb,showpacs,10pt,superscriptaddress,preprintnumbers]{revtex4-1}
\pdfoutput=1

\usepackage{natbib}
\usepackage{amsmath,amssymb}
\usepackage{wasysym}
\usepackage[utf8]{inputenc}
\usepackage{hyperref}
\usepackage{graphicx}
\usepackage{epstopdf}
\usepackage{slashed}
\begin{document}

\title{Effective field theory interpretation of searches for dark matter annihilation in the Sun with the IceCube Neutrino Observatory}

\author{Jan Blumenthal}
\email{blumenthal@physik.rwth-aachen.de}
\affiliation{III. Physikalisches Institut B, RWTH Aachen University, 52056 Aachen, Germany}

\author{Pavel Gretskov}
\email{gretskov@physik.rwth-aachen.de}
\affiliation{III. Physikalisches Institut B, RWTH Aachen University, 52056 Aachen, Germany}

\author{Michael Kr\"amer}
\email{mkraemer@physik.rwth-aachen.de}
\affiliation{Institute for Theoretical Particle Physics and Cosmology, RWTH Aachen University, 52056 Aachen, Germany\\ and SLAC National Accelerator Laboratory, Stanford University, Stanford, CA 94025, USA}

\author{Christopher Wiebusch}
\email{wiebusch@physik.rwth-aachen.de}
\affiliation{III. Physikalisches Institut B, RWTH Aachen University, 52056 Aachen, Germany}

\date{\today}

\preprint{SLAC-PUB-16136}

\begin{abstract}

We present a model-independent interpretation of searches for dark matter annihilation in the Sun using an effective field theory approach. We identify a set of effective operators contributing to spin-dependent scattering of dark matter with protons in the non-relativistic limit and explore simple new physics models which would give rise to such operators. Using the limits on the spin-dependent scattering cross-section set by the IceCube collaboration in their search for dark matter annihilation in the Sun, we derive limits on effective couplings and corresponding masses of mediating particles. We show that the effective field theory interpretation of the IceCube searches provides constraints on dark matter complementary to those from relic density observations and searches at the LHC. Finally, we discuss the impact of astrophysical uncertainties on our results. 

\end{abstract}


\maketitle


\section{Introduction}
\label{sec:intro}

Since its first discovery by Zwicky in the 1930s \cite{Zwicky1937}, there has been an ever-growing list of observational evidence for the existence of dark matter. These observations led to the understanding that presumably every galaxy, including our own Milky Way, is surrounded by a halo of dark matter \cite{Begeman1991, Hayashi2004, Navarro1996, Moore1999, Burkert1995}. It became clear, that in order to consolidate this evidence with the Standard Model of particle physics (SM), which does not include a suitable candidate for dark matter, the SM would have to be extended. Some of the extensions, like supersymmetry (SUSY) or axions, deal with more general questions in particle physics and naturally provide dark matter candidate particles, while other models take a more \textit{ad hoc} approach to dark matter (for an extensive list see \cite{Bertone2005}). 

One popular idea is that dark matter could be a stable weakly interacting massive particle (WIMP) with masses near the TeV-scale and weak interaction strengths. This allows to explain the present dark matter density in the universe by the freeze-out a particle, which in the early universe was in thermal equilibrium with the rest of the primordial plasma. This so-called ``WIMP miracle'' combined with the fact that models like supersymmetry predict a suitable candidate for WIMPs are the two main reasons for its popularity. While we will try to provide a mostly model-independent approach to dark matter, we will nevertheless focus on the WIMP scenario and therefore from now on use these two expressions synonymously.

There is a considerable experimental effort to detect dark matter in particle physics processes. This includes direct and indirect detection, as well as searches at particle accelerators. Direct detection relies on measuring the nuclear recoil from a WIMP scattering off a nucleus in a very low-background environment, consisting of either liquid Xenon (XENON \cite{Aprile2011, Aprile2009}, LUX  \cite{Akerib2014}) or some solid target material (CDMS \cite{Akerib2003}, EDELWEISS \cite{Lemrani2006}, CRESST \cite{Angloher2012}). Indirect searches aim to find various products of dark matter annihilation, such as $\gamma$-rays, neutrinos or antiparticles, from sources inside and outside our galaxy. This includes satellite-based experiments (Fermi\-LAT \cite{Ackermann2011, Ackermann2012}, PAMELA \cite{Picozza2007}, AMS \cite{Barao2004, Aguilar2013}), ground-based experiments (H.E.S.S. \cite{Hinton2004}, VERITAS \cite{Weekes2002}), and subsurface neutrino telescopes (BUST \cite{Baksan:ICRC1979}, Baikal Neutrino Project \cite{Avrorin2011}, Super-Kamiokande \cite{Abe2011}, ANTARES \cite{Ageron2011}, IceCube \cite{Achterberg2006}).
Finally, searches with the experiments ATLAS and CMS at the Large Hadron Collider (LHC) aim to produce dark matter in collisions of SM particles and then infer its presence from missing energy in the final state \cite{Aad2013, CMS2014}.
While many indirect searches target annihilation products of WIMPs in galactic halos (\textit{e.g.}\ the Milky Way halo) or the Galactic Center, neutrino telescopes are also sensitive to annihilation in the center of massive celestial
bodies, in particular the Sun. In 2013 the IceCube Collaboration has published the currently most stringent exclusion limits for WIMP annihilations in the Sun \cite{Aartsen2013}. 
In this paper we will interpret these results in an effective field theory approach and compare these with searches at the LHC.

The paper is structured as follows. In Section~\ref{sec:capture} we review the physics of WIMP capture and annihilation in the Sun. The search for dark matter annihilation in the Sun with IceCube is presented in Section~\ref{sec:icresults}. We shall focus on the interpretation of the search in an effective theory with spin-dependent interactions between WIMPs and quarks, as discussed in Section~\ref{sec:eftanalysis}. Simple models of physics beyond the SM which would lead to such effective interactions are introduced in Section~\ref{sec:BSMWIMP}. To obtain limits on the effective couplings of WIMPs with matter from the annihilation in the Sun, we need to calculate the capture rate of WIMPs in the Sun. The capture rate is given in terms of the WIMP-nucleus scattering cross-sections presented in Section~\ref{sec:cross-section}. In Section~\ref{sec:cross-section} we also present a calculation of the dark matter relic density within the effective field theory, which provides complementary information on the allowed WIMP masses and couplings. Our results are presented in Section~\ref{sec:results}, where we show the limits on the effective WIMP-quark couplings and a comparison with constraints from the dark matter relic density and searches at the LHC. The astrophysical uncertainties from the local WIMP density and velocity distribution are discussed in Section~\ref{sec:uncertainties}. We conclude in Section~\ref{sec:conclusions}.


\section{WIMP capture and annihilation in the Sun}\label{sec:capture}

The idea that dark matter from the halo could be accumulated by celestial bodies passing through it was put forth in the 1980s by a variety of physicists (see \cite{Gould1992} and references therein). 
Since then it has been established as one of the standard methods for indirect detection of dark matter and was employed by several Earth-bound experiments \cite{Adrian2013, Kappl2011,Abbasi2011a, Boliev2013, Avrorin2015}.

The number of dark matter particles in the Sun, $N$, is governed by the Riccati differential equation \cite{Jungman1996}
\begin{equation}
\label{diffeqcap}
\dot{N} = C_{\astrosun} - C_A N^2 - C_E N,
\end{equation}
where $\dot{N}$ denotes the derivative with respect to time, $C_{\astrosun}$ is the rate at which new dark matter particles are captured, $C_A N^2 = 2 \Gamma_A$ is twice the rate at which dark matter annihilates, and $C_E N$ accounts for the escape of particles due to hard elastic scattering, also called evaporation.
The parameter $C_A$ is responsible for the depletion of dark matter particles through self-annihilation. It is given by $C_A = \langle \sigma_A v \rangle/V_\text{eff}$, where $\langle \sigma_A v \rangle$ is the velocity-averaged annihilation cross-section and $V_\text{eff}$ is the effective volume of the WIMP core \cite{Jungman1996}. Note that $\langle \sigma_A v \rangle$ is calculated in the limit of zero relative velocity, since the WIMPs in the Sun are highly non-relativistic. The last term in Eq.~(\ref{diffeqcap}), $C_E N$, was shown to be negligible in the case of the Sun for WIMPs with $m_\chi \gtrsim 10\, \text{GeV}$~ \cite{Griest1987}. 
The capture rate $C_{\astrosun}$ depends on the WIMP density and velocity distribution and on the elastic WIMP-nucleon scattering cross-section \cite{Gould1987}. 
For the numerical computation of the capture rate we use the computer code \texttt{DarkSUSY} \cite{Gondolo2004}. 

It is crucial for our analysis that the WIMP-nucleon scattering cross-section has a spin-dependent (SD) and a spin-independent (SI) part. Due to coherent scattering off all nucleons in an atom, SI scattering has a quadratic dependence on the mass number $A$, $\sigma_{\rm SI} \propto A^2$, which leads to strong enhancement for heavy elements.  SD scattering on the other hand depends on the total nuclear angular momentum $\sigma_{\rm SD} \propto J_N$ and is sub-dominant when the target material does not contain a large abundance of elements with unpaired spin. For the Sun, however, there is a large abundance of target material with non-zero nuclear angular momentum in the form of hydrogen, and thus the SD scattering contributes significantly to the total scattering rate and WIMP capture. The capture rate $C_{\astrosun}$  also depends on the velocity distribution of WIMPs in the halo which is usually assumed to be a Maxwell-Boltzmann distribution. However, recently other velocity distributions have been studied and shown to have significant effects on solar capture rates \cite{Choi2013}. We will discuss the impact of the choice of the velocity distribution on our results in Section~\ref{sec:uncertainties}.

Neglecting the evaporation term, the solution of Eq.~(\ref{diffeqcap}) is 
\begin{equation}\label{eq:tanhsq}
\Gamma_A = \frac{C_{A} N^2}{2} = \frac{1}{2} C_{\astrosun} \tanh^2 (\sqrt{C_{\astrosun} C_A} \, t).
\end{equation}
For large times $t$ and correspondingly large values of $\sqrt{C_{\astrosun} C_A} \, t \gg 1$,  the $\tanh$-term becomes $\simeq 1$, and the annihilation rate depends only on the capture rate, but not on the annihilation cross section:
\begin{equation}
\label{equil}
\Gamma_A = \frac{1}{2} C_{\astrosun} \equiv \frac{1}{2} (K_{\rm SI} \sigma_{\rm SI} + K_{\rm SD} \sigma_{\rm SD}).
\end{equation}
Here, $K_{\rm SI}$ and $K_{\rm SD}$ are capture efficiencies for the SI and SD parts of the scattering. For $C_{\astrosun} C_A \, t \gg 1$, WIMP annihilation and capture are in equilibrium, $C_{\astrosun} = 2 \Gamma_A = C_A N^2$, and thus $\dot{N} = 0$. 

Through a measurement of the neutrino flux, neutrino telescopes are sensitive to the WIMP annihilation rate. If WIMP capture and annihilation are in equilibrium, the annihilation rate determines the capture rate, $C_{\astrosun} = 2 \Gamma_A$, which in turn provides information on the elastic WIMP scattering cross sections $\sigma_{\rm SI}$ and $\sigma_{\rm SD}$ probed in direct detection experiments.  

Assuming that the Sun has been collecting WIMPs during its whole lifetime, 
 $t = t_{\astrosun} \simeq 1.5 \times 10^{17} \, \text{s}$, results
in the approximate expression \cite{Jungman1996}
\begin{widetext}
\begin{equation}
\label{eq:equil}
\sqrt{C_{\astrosun} C_A} \, t_{\astrosun} \simeq 330 \left( \frac{C_{\astrosun}}{\text{s}^{-1}} \right)^{1/2} \left( \frac{\langle \sigma_A v \rangle}{\text{cm}^3 \, \text{s}^{-1}} \right)^{1/2} \left( \frac{m_\chi}{10 \, \text{GeV}} \right)^{3/4}.
\end{equation}
\end{widetext}
With WIMP scattering cross sections, and thus capture rates, at the level of the current experimental upper limits, and annihilation cross sections $\sim 10^{-26} \, \text{cm}^3 \, \text{s}^{-1}$, as needed to explain the dark matter relic density, one finds $\sqrt{C_{\astrosun} C_A} \, t_{\astrosun}  \gg 1$, \textit{i.e.}\ the assumption of equilibrium for WIMP annihilation and capture in the Sun is in general well satisfied. Nevertheless, as we will discuss in Section \ref{sec:results}, in some astrophysical scenarios equilibrium may not hold, and the non-equilibrium effects have to be taken into account.


\section{Search for dark matter annihilation in the Sun with the IceCube neutrino Observatory}
\label{sec:icresults}

The search for WIMP annihilations in the Sun with the IceCube Neutrino Observatory \cite{Aartsen2013} is based on the detection
of muon-neutrinos by charged current interactions. The analysis uses data collected from June 2010 to May 2011 with the IceCube detector in its 79-string configuration including 6 strings of the low-energy sub-detector DeepCore \cite{Collaboration:2011ym}.
The detector measures Cherenkov light emitted by secondary muon tracks and reconstructs their directions. The angles $\Psi$ between the selected muon neutrino events with respect to the direction to the Sun are used to test for an enhanced flux from that direction relative to the background dominated by atmospheric neutrinos. With this method IceCube is able to detect neutrinos with energies above a threshold energy of $\sim 100 \, \text{GeV}$ with the main detector. The energy threshold can become as low as $\sim 10 \, \text{GeV}$ when including DeepCore into the analysis. 

The signal prediction was obtained from simulations using \texttt{WimpSim} \cite{Blennow2008} with two extreme benchmark scenarios. In these scenarios two channels for WIMP annihilation are chosen according to the type of neutrino spectrum they would produce at Earth and assigned a $100 \%$ branching ratio. As a conservative scenario one assumes, that all WIMPs annihilate into pairs of b and $\bar{\,\rm b}$ quarks. Since the b-quarks are able to hadronize into B-mesons in the interior of the Sun, they lose a large portion of their initial energy before producing neutrinos \cite{Kamionkowski1991}. Therefore the resulting neutrino spectrum would be steeply falling with increasing neutrino energy (``soft''). The other extreme would be if all WIMPs annihilated into pairs of electroweak W-bosons, or
pairs of $\tau^+$ and $\tau^-$ if the invariant mass is below the production threshold of W-bosons. Due to the short lifetime, these particles  
do not lose a significant part of their energy before decaying into neutrinos. The resulting neutrino spectrum would produce more highly energetic neutrinos at Earth (``hard''). From this, the probability distributions for the solid angle between the direction of the Sun and of the reconstructed neutrinos are obtained for signal events.

No significant excess over the background expectation was found and a $90 \%$ confidence level upper limit on the number of muon neutrino signal events, $\mu_s^{90}$, was derived as a function of the assumed WIMP mass. From the limit on the number of neutrino signal events $\mu_s^{90}$ a limit on the annihilation rate in the Sun $\Gamma_A$ was derived using \texttt{DarkSUSY} \cite{Gondolo2004}. 
Assuming equilibrium, the limit on $\Gamma_A$ was then converted to upper limits on SI and SD WIMP-proton scattering cross-sections according to Eq.~ (\ref{equil}). In order to do so, one assumes that capture is either exclusively SI or exclusively SD. Using \texttt{DarkSUSY} one can then calculate the capture efficiencies $K_{\rm SI}$ and  $K_{\rm SD}$ in Eq.~(\ref{equil}) and from that arrive at values for the scattering cross sections $\sigma_{\rm SI}$ and $\sigma_{\rm SD}$, respectively. Note that since the annihilation rate, in general, depends on the sum of SI and SD scattering cross sections, setting either one to zero yields a conservative upper limit on the other one. In the following, we will focus on the limits for the SD scattering cross-section from neutrino fluxes, as these are more stringent than limits produced by direct detection experiments.  


\section{Effective Operator analysis}
\label{sec:eftanalysis}

Assuming that the interaction between WIMP dark matter and the Standard Model is mediated by particles with masses $M$ much larger than that of the WIMP, $M \gg m_\chi$, 
the WIMP-SM interaction can be described by an effective Lagrangian organised in inverse powers of the scale of new physics $\Lambda \sim M$\,\cite{Beltran:2008xg}: 
\begin{equation}
 \mathcal{L} = \mathcal{L}_\text{SM} + \sum_{n > 4} \frac{f^{(n)}}{\Lambda^{n-4}} \mathcal{O}^{(n)}. 
\end{equation}
Here, $f^{(n)}$ are dimensionless constants, and $\mathcal{O}^{(n)}$ are operators of mass-dimension $n$ describing the interactions between the WIMP and the SM particles. Effective field theory interpretations of direct, indirect and collider searches for dark matter have been studied widely in the literature, see \textit{e.g.}\,\cite{Beltran:2008xg, Birkedal:2004xn, Cao:2009uw, Beltran:2010ww, Bai:2010hh, Goodman:2010ku, Rajaraman:2011wf, Fox:2011pm, Cheung:2012gi, Frandsen:2012rk, Dreiner2013, Chae:2012bq, Rajaraman:2012fu, Dreiner:2013vla, Haisch:2013fla, Krauss:2013wfa, Berger:2014sqa}. 

The effective field theory approach is valid only as long as the typical energy scale of an interaction is smaller than the scale that characterises the UV-completion of the new physics, $E \lesssim \Lambda\sim M$. We will discuss the validity of the effective field theory approach in the context of the IceCube search for dark matter annihilation in the Sun in Section \ref{sec:results}. 

We assume a minimal dark matter model with a $Z_2$ symmetry and focus on spin-dependent interactions 
for which 
the search for neutrinos from WIMP annihilation in the Sun is usually 
more sensitive  than direct detection. This drastically limits the number of effective operators, as we will show below~\cite{Agrawal2010a, Kumar:2013iva, Pavel:Thesis:2014}. 

The matrix elements for the scattering of a WIMP on a nucleus $N$ involve the expectation values of quark bilinears, $\langle N | \bar{q} \Gamma q | N \rangle$, where  $\Gamma \in \{\mathbb{I}, \gamma^5, \gamma^\mu, \gamma^\mu\gamma^5,\sigma^{\mu\nu} \}$.
Of these, only the axial-vector and tensor interactions, $\langle N | \bar{q} \gamma^\mu \gamma^5 q | N \rangle$ and $\langle N | \bar{q} \sigma^{\mu\nu} q | N \rangle$, respectively, contribute to spin-dependent scattering in the non-relativistic limit. 

We now proceed to examine the possible higher-dimensional  operators which involve axial-vector and tensor quark bilinears, discussing scalar, fermion and vector dark matter in turn. 

Scalar dark matter, $\phi$, can couple to the axial-vector and tensor quark bilinears through operators which involve derivatives, like for example $\phi \partial_\mu \phi \bar{q}\gamma^\mu\gamma^5 q$. However, for both the temporal and spatial components of $\mu$ the matrix elements of such operators are velocity suppressed and thus vanish in the non-relativistic limit. Note that no such suppression exists for the coupling of scalar dark matter to the vector quark bilinear $\bar{q} \gamma^\mu q$, so that non-relativistic nucleon scalar-WIMP scattering is in general spin-independent. 

There are three ways to couple fermionic dark matter, denoted by a spinor field $\chi$, to the axial-vector and tensor quark bilinears: $\bar{\chi} \gamma_\mu \chi \bar{q} \gamma^\mu \gamma^5 q$,  
$ \bar{\chi} \gamma_\mu \gamma^5 \chi \bar{q} \gamma^\mu \gamma^5 q$ and 
 $\bar{\chi} \sigma_{\mu \nu} \chi \bar{q} \sigma^{\mu \nu} q$.  The first term again is velocity-suppressed, while the second and third term indeed contribute to spin-dependent WIMP-nucleon scattering in the non-relativistic limit. If $\chi$ is a Majorana fermion, then $\bar{\chi} \sigma_{\mu \nu} \chi$ identically vanishes, leaving $ \bar{\chi} \gamma_\mu \gamma^5 \chi \bar{q} \gamma^\mu \gamma^5 q$ as the only operator contributing to non-relativistic Majorana-WIMP scattering. 

The analysis of vector dark matter, denoted by $B^\mu$ is slightly more involved, see \cite{Agrawal2010a, Kumar:2013iva, Pavel:Thesis:2014} for details. It is straightforward to verify that all combinations of vector fields and a tensor quark bilinear either vanish identically or are velocity suppressed. Similarly, by explicit calculation, and by making use of the equation of motion for a vector field, one can show that operators of the type $B^\nu (\partial_\mu B_\nu)\bar q \gamma^\mu \gamma^5 q$ or $B_\mu (\partial^\nu B_\nu) \bar q \gamma^\mu \gamma^5 q$ are velocity suppressed or vanish identically. However, the operator  $\epsilon^{\mu \nu \rho \sigma} B_\mu (\partial_\nu B_\rho) \bar q \gamma_\sigma \gamma^5 q$ can be shown to contribute to vector dark mater spin-dependent scattering in the non-relativistic limit. For complex dark matter, the corresponding operator $\epsilon^{\mu \nu \rho \sigma} B_\mu^\dagger (\partial_\nu B_\rho) \bar q \gamma_\sigma \gamma^5 q$ and the operator $(B_\mu^\dagger B_\nu - B_\nu^\dagger B_\mu)\bar{q}\sigma^{\mu\nu}q$ also contribute to spin-dependent scattering. However, they are in general accompanied by the operators $B^\nu (\partial_\mu B_\nu^\dagger)\bar q \gamma^\mu \gamma^5 q$ and $B_\mu^\dagger B^\mu \bar{q} q$, respectively, which contribute to spin-independent scattering. Thus, in order for spin-dependent interactions to dominate, one would need to suppress those operators,  for example by imposing an additional discrete symmetry. We will thus focus on real vector dark matter in this analysis. 

To summarise this Section, we found that there are four types of effective operators with mass-dimension 5 or 6 giving rise to spin-dependent WIMP-nucleon scattering in the non-relativistic limit. 
While the non-relativistic scattering of scalar WIMPs is spin-independent in general, there are two operators which lead to spin-dependent interactions for fermionic dark matter:
\begin{equation}
\begin{aligned}
\label{eq:fermiwimp}
 \mathcal{O}^{(6)}_{AA} &= \bar{\chi} \gamma_\mu \gamma^5 \chi \bar{q} \gamma^\mu \gamma^5 q,  \\ 
 \mathcal{O}^{(6)}_{TT} &= \bar{\chi} \sigma_{\mu \nu} \chi \bar{q} \sigma^{\mu \nu} q.
\end{aligned}
\end{equation}
Note that for Majorana fermions, the tensor operator $\mathcal{O}^{(6)}_{TT}$ vanishes identically. Moreover, there are 
two kinds of operators which result in spin-dependent interactions for vector dark matter: 
\begin{equation}
\label{eq:vectorwimp}
\begin{aligned}
 \mathcal{O}^{(6)}_{VB} &= \epsilon^{\mu \nu \rho \sigma} B_\mu^{(\dagger)} (\partial_\nu B_\rho) \bar q \gamma_\sigma \gamma^5 q, \\
 \mathcal{O}^{(5)}_{VB} &= (B_\mu^\dagger B_\nu - B_\nu^\dagger B_\mu)\bar{q}\sigma^{\mu\nu}q.
\end{aligned}
\end{equation}
In general, the operators listed in Eq.~(\ref{eq:fermiwimp}) and Eq.(\ref{eq:vectorwimp}) are accompanied by operators which induce spin-independent interactions. However, for Majorana fermions and real vector dark matter with chiral couplings (see Section~\ref{sec:BSMWIMP}), 
the spin-independent operators vanish, and one can thus expect predominantly spin-dependent interactions.  


\section{Simplified models for spin-dependent interactions}
\label{sec:BSMWIMP}
It is instructive to explore possible extensions of the Standard Model which would give rise to the effective operators for spin-dependent WIMP-quark interactions, Eqs.~(\ref{eq:fermiwimp}) and (\ref{eq:vectorwimp}), at tree level. We are not attempting to explore complex scenarios like supersymmetric theories, but rather consider simplified models with a minimal field content. Such models have been studied in some detail recently in the context of collider physics, see \textit{e.g.} \cite{Abdallah:2014hon, Malik:2014ggr} and references therein. In this section we shall construct renormalizable Lagrangians containing a fermionic or vector WIMP and one scalar, fermion or vector boson mediator, and explore the possible spin-dependent interactions between the WIMP and the SM quarks. 

\subsection{Fermionic WIMP}

\begin{figure}
\includegraphics[width=0.23\textwidth]{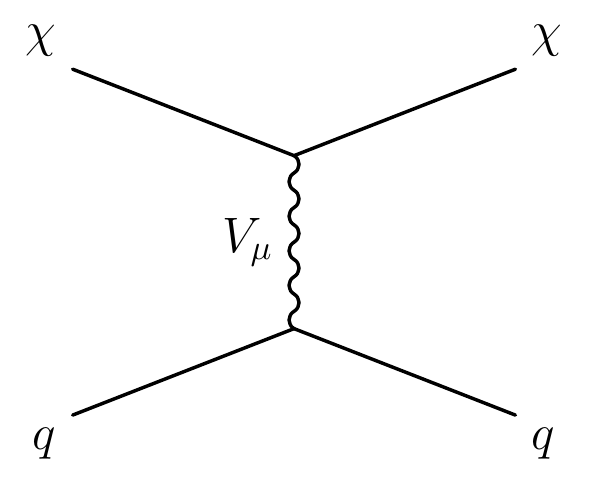}
\includegraphics[width=0.23\textwidth]{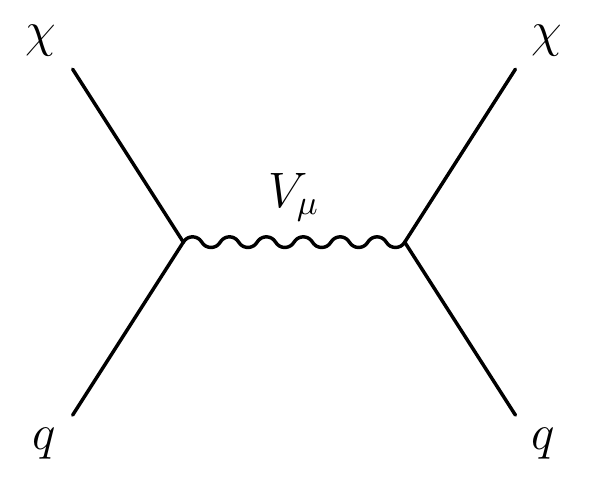}
\includegraphics[width=0.23\textwidth]{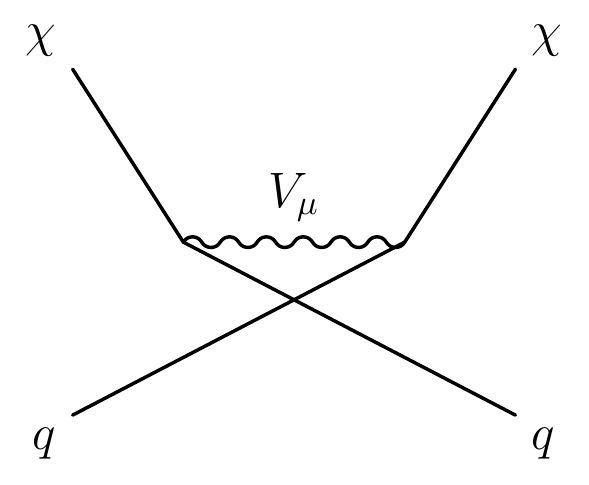}
\caption{Feynman diagrams for the scattering of a fermionic WIMP off quarks through the exchange of a vector particle: $t$-, $s$- and $u$-channel contributions (upper left, upper right and lower panel, respectively).}
\label{fig:fermion-feynman}
\end{figure}

We first consider fermionic dark matter, with interactions mediated by heavy vector or scalar particles. The tree-level scattering processes include $t$, $s$ and $u$-channel contributions as shown in FIG.~\ref{fig:fermion-feynman} for a vector boson mediator (and analogously for scalar mediators) 

The simplified model Lagrangian describing WIMP-quark scattering through the exchange of a vector mediator, $V_\mu$,  in the $t$-channel is given by \cite{Dreiner2013, Agrawal2010a}
\begin{eqnarray}\label{eq:lagrangian_FV}
 \mathcal{L} = &-& \frac{1}{4} F^{\mu \nu} F_{\mu \nu} + \frac{1}{2} m^2_V V_\mu V^\mu - \bar q \gamma^\mu (g_L P_L + g_R P_R) q V_\mu \nonumber \\
 &-& \bar \chi \gamma^\mu (g'_L P_L + g'_R P_R) \chi V_\mu,
\end{eqnarray}
where $F^{\mu \nu} = \partial^\mu V^\nu - \partial^\nu V^\mu$ is the field strength tensor and $m_V$ is the mass of the vector boson, $P_{L,R} = (1 \pm \gamma^5)/2$ are the left- and right-handed projection operators, and $g^{}_{L,R}, g'_{L,R}$ are dimensionless coupling constants. Note that we have omitted the kinetic and mass terms for the fermions, since they do not play a role in the calculation of the effective operators. Integrating out the vector field, we obtain the effective Lagrangian of the WIMP-quark interaction: 
\begin{equation}
 \mathcal{L}_\text{eff} = - \frac{1}{m^2_V} \bar q \gamma^\mu (g_V - g_A \gamma^5) q \bar \chi \gamma_\mu (g'_V - g'_A \gamma^5) \chi,
\end{equation}
with the coupling strengths $g_V = (g_L + g_R)/2$ and $g_A = (g_L - g_R)/2$, and correspondingly for  $g'_{V,A}$. The Lagrangian in Eq.~(\ref{eq:lagrangian_FV}) thus gives rise 
to four effective operators. As we have shown in Section~\ref{sec:eftanalysis}, the two terms coupling the axial-vector quark bilinear to the vector WIMP bilinear and vice versa vanish 
in the non-relativistic limit. The remaining two terms contain a pure vector and a pure axial-vector coupling:
\begin{equation}
\label{eq:axialtchannel}
\frac{g_V g'_V }{m^2_V} \bar q \gamma^\mu q \chi \gamma_\mu \chi \quad \mbox{and}\quad 
\frac{g_A g'_A }{m^2_V} \bar q \gamma^\mu \gamma^5 q \chi \gamma_\mu \gamma^5 \chi.
\end{equation}
The first term is indeed present in the non-relativistic limit and contributes to spin-independent scattering, while the second term contains the operator $\mathcal{O}^{(6)}_{AA}$, which we have shown to give rise to spin-dependent scattering. Note, that for a Majorana WIMP  the first term vanishes due to charge conjugation symmetry, and scattering would be predominantly spin-dependent.

Let us now consider the case of $s$- and $u$-channel exchange of a vector mediator. The corresponding simplified model Lagrangian is \cite{Dreiner2013, Agrawal2010a}
\begin{eqnarray}
 \mathcal{L} =&-& \frac{1}{2} F^{\mu \nu} F^\dagger_{\mu \nu} + m^2_V V^\dagger_\mu V^\mu - \bar q \gamma^\mu (g_V - g_A \gamma^5) \chi V^\dagger_\mu \nonumber \\
 &-& \bar \chi \gamma^\mu (g^*_V - g^*_A \gamma^5) q V_\mu,
\end{eqnarray}
where we have already inserted the explicit expressions of the left- and right-handed projection operators to write the fermion currents in the standard $V-A$ form. The corresponding effective Lagrangian for the WIMP-quark interactions reads 
\begin{equation}
 \mathcal{L}_\text{eff} = - \frac{1}{m^2_V} \bar q \gamma^\mu (g_V - g_A \gamma^5) \chi \bar q \gamma_\mu (g^*_V - g^*_A \gamma^5) \chi.
\end{equation}
We can use Fierz identities \cite{Nieves2004} to rearrange the terms and arrive at
\begin{eqnarray}
 \mathcal{L}_\text{eff} = &-& \frac{1}{m^2_V} \Bigl[ (|g_V|^2 - |g_A|^2) \bar q q \bar \chi \chi \nonumber \\ 
 &-& \frac{1}{2} (|g_V|^2 + |g_A|^2) \bar q \gamma^\mu q \bar \chi \gamma_\mu \chi \nonumber \\  
 &-& \frac{1}{2} (|g_V|^2 +  |g_A|^2) \bar q \gamma^\mu \gamma^5 q \bar \chi \gamma_\mu \gamma^5 \chi \Bigr],
\end{eqnarray}
where we have omitted terms not contributing in the non-relativistic limit. The effective Lagrangian includes the axial-vector operator $\mathcal{O}^{(6)}_{AA}$, but also operators with scalar and vector coupling. This simplified model will therefore give rise to both spin-dependent and independent scattering. If we assume that the WIMP is a Majorana fermion, the vector operator would vanish.  The scalar operator, however, is even under charge conjugation and would therefore still contribute to spin-independent scattering. Therefore, in order for spin-dependent scattering to dominate, we would have to further assume that $g_V = \pm g_A$ so that the pre-factor of the scalar term vanishes. This corresponds to the so-called chiral limit where either the left-handed coupling $g_L$ or the right-handed coupling $g_R$ is zero, \textit{i.e.}\ only the left- or right-handed components of the WIMP couple to the mediating vector boson. 

Note that for $s$- and $u$-channel interactions the vector boson must carry color and electromagnetic charge. Furthermore, the Lagrangian does not exhibit a $Z_2$-symmetry in the WIMP sector,  and the WIMP can thus decay into $V_\mu$ and a lighter SM particle. In order to assure that the WIMP can account for the present day relic abundance, the WIMP would have to be lighter than the mediator so that the decay is kinematically inaccessible. 

Interactions of a fermionic WIMP with quarks can also be mediated by a scalar particle $\phi$, with $t$-, $s$- and $u$-channel contributions as for vector mediators. The $t$-channel exchange, 
described by the  Lagrangian 
\begin{eqnarray}
 \mathcal{L} =&& \frac{1}{2} (\partial_\mu \phi)(\partial^\mu \phi) - \frac{1}{2} m^2_\phi \phi^2 - \bar q (g_L P_L + g_R P_R) q \phi \nonumber \\
 &&- \bar \chi (g'_l P_L + g'_r P_R) \chi \phi, 
\end{eqnarray}
only leads to scalar operators, which do not contribute to spin-dependent scattering as argued in Section~\ref{sec:eftanalysis}. 

However, $s$- and $u$-channel interactions with the Lagrangian 
\begin{eqnarray}
\mathcal{L} &=& (\partial_\mu \phi^\dagger)(\partial^\mu \phi) - m^2_\phi \phi^\dagger \phi - \bar q (g_S + g_P \gamma^5) \chi \phi^\dagger \nonumber \\
&&- \bar \chi (g^*_S + g^*_P \gamma^5) q \phi
\end{eqnarray}
lead to an effective interactions of the type 
\begin{equation}
 \mathcal{L}_\text{eff} = - \frac{1}{m^2_\phi} \bar q (g_S - g_P \gamma^5) \chi \bar q (g^*_V + g^*_A \gamma^5) \chi, 
\end{equation}
where $g_S=(g_L + g_R)/2$ and $g_P=(g_R - g_L)/2$. Using Fierz identities to rearrange the spinor bilinears and neglecting terms which vanish in the non-relativistic limit we arrive at 
\begin{eqnarray}
 \mathcal{L}_\text{eff} =&-& \frac{1}{m^2_\phi} \Bigl[ \frac{1}{4} (|g_S|^2 - |g_P|^2) \bar q q \bar \chi \chi \nonumber \\
 &+& \frac{1}{4} (|g_S|^2 + |g_P|^2) \bar q \gamma^\mu q \bar \chi \gamma_\mu \chi \nonumber \\  
 &-& \frac{1}{4} (|g_S|^2 + |g_P|^2) \bar q \gamma^\mu \gamma^5 q \bar \chi \gamma_\mu \gamma^5 \chi \nonumber \\
 &+& \frac{1}{8} (|g_S|^2 - |g_P|^2) \bar q \sigma^{\mu \nu} q \bar \chi \sigma_{\mu \nu} \chi \Bigr].
\end{eqnarray}
We see that both operators found in the last section $\mathcal{O}^{(6)}_{AA}$ and $\mathcal{O}^{(6)}_{TT}$ arise in this model. However, they are accompanied by the scalar and vector operators which contribute to spin-independent scattering. As in the case for a vector mediator, additional theoretical assumptions have to be made to construct a model where spin-dependent scattering dominates. The assumption that the WIMP is a Majorana fermion eliminates the vector and tensor operators, leaving the scalar operator as the only spin-independent contribution. For it to vanish, the scalar and pseudoscalar couplings must be of equal strength $g_S = \pm g_P$, corresponding to the chiral limit of only left- or right-handed WIMPs coupling to the scalar mediator. Note, that  the scalar boson mediating $s$- and $u$-channel exchange is charged under the electromagnetic and strong force. Also, the WIMP is no longer protected by a $Z_2$-symmetry, and will decay unless it is lighter than the mediator particle.

\subsection{Vector Boson WIMP}

\begin{figure}
\includegraphics[width=0.23\textwidth]{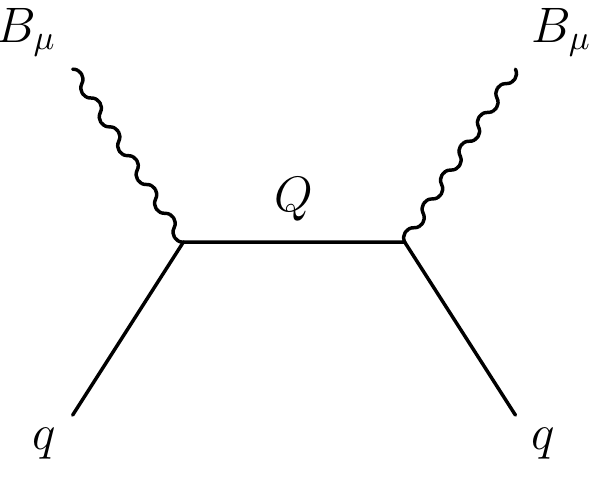}
\includegraphics[width=0.23\textwidth]{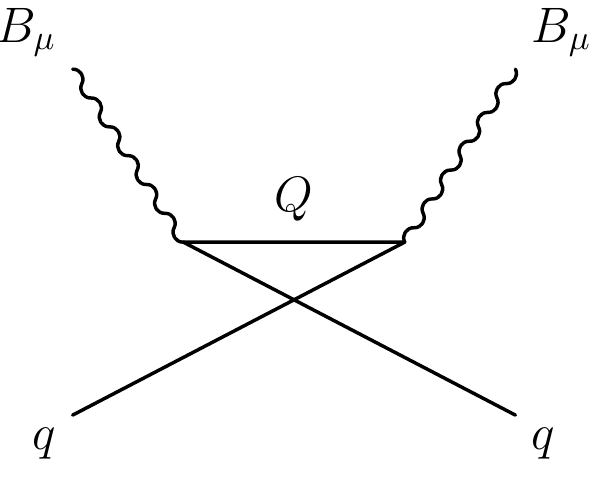}
\caption{Feynman diagrams for the scattering of a vector WIMP off quarks through the exchange of a fermion: $s$- (left) and $u$-channel (right) contributions.}
\label{fig:vector-feynman}
\end{figure}

The interaction of a vector boson WIMP with SM quarks can be mediated by $t$-channel exchange of scalar and vector particles. It is straightforward to show \cite{Agrawal2010a} that they do not 
contribute to spin-dependent scattering. We will therefore not consider them further and instead focus on the $s$- and $u$-channel exchange of fermionic mediators, FIG.~\ref{fig:vector-feynman}, 
with the Lagrangian \cite{Agrawal2010a}
\begin{eqnarray}
\mathcal{L} =&& \bar Q (i \slashed{\partial} - m_Q) Q - \bar q \gamma^\mu (g_V - g_A \gamma^5) Q B_\mu \nonumber \\
&&- \bar Q \gamma_\mu (g^*_V - g^*_A \gamma^5) q B_\mu.
\end{eqnarray}
As mentioned in Section~\ref{sec:eftanalysis} we focus on real vector dark matter, denoted by $B_\mu$. Integrating out the fermion mediator we 
obtain the effective Lagrangian
\begin{eqnarray}
\label{eq:leffvect}
 \mathcal{L}_\text{eff} &=& \frac{1}{m_Q} \Bigl[ \bar q \gamma^\mu (g_V - g_A \gamma^5) \gamma^\nu (g^*_V - g^*_A \gamma^5) q B_\nu B_\mu \Bigr] \\
 &&+ \frac{i}{m^2_Q} \Bigl[ \bar q \gamma^\mu (g_V - g_A \gamma^5) \gamma^\alpha \gamma^\nu (g^*_V - g^*_A \gamma^5) \partial_\alpha(q B_\nu B_\mu) \Bigr] \nonumber.
\end{eqnarray}
Expanding the two terms, and neglecting contributions that are either velocity-suppressed or suppressed by powers of ratios of SM quark and WIMP masses, $m_q/m_B$, we find that only two effective operators remain: 
\begin{eqnarray}
\mathcal{L}_\text{eff} &=& \frac{1}{m_Q} (|g_V|^2 - |g_A|^2) \bar q q B^\mu B_\mu \nonumber \\
&&- \frac{1}{m^2_Q} (|g_V|^2 + |g_A|^2) \epsilon^{\mu \nu \rho \sigma} \bar q \gamma_\rho \gamma^5 q B_\nu \partial_\sigma B_\mu.
\end{eqnarray}
The second term corresponds to the operator $\mathcal{O}^{(6)}_{VB}$ which, as shown in Section~\ref{sec:eftanalysis}, contributes to spin-dependent scattering. In order for this operator to dominate, we again have to assume the chiral limit $g_V = \pm g_A$, such that the pre-factor of the scalar quark bilinear vanishes.\newline

Concluding this section, we found that there are various simplified models which create effective operators with spin-dependent interactions. For fermionic WIMPs, both spin-dependent and spin-independent interactions arise in general from the exchange of scalar or vector mediators. For models with Majorana fermion dark matter and $t$-channel exchange of neutral vector mediators, the spin-independent operators are suppressed, and the scattering cross-section is expected to be predominantly spin-dependent. For the exchange of charged mediators in the $s$- and $u$-channel, one has to assume chiral couplings to suppress the spin-independent interactions. Spin-dependent interactions also arise for real vector WIMPs and a charged fermionic mediator. However, also in this case we have to assume chiral couplings to suppress the spin-independent interactions. 


\section{Cross sections and relic density}\label{sec:cross-section}

To interpret the IceCube measurement of the neutrino flux in terms of the effective field theory, we have to determine the capture rate and thus the elastic scattering cross section of dark matter particles in the Sun, see Eq.~(\ref{equil}). Recall that IceCube places upper limits on the annihilation rate, which is determined by the capture rate if dark matter capture and annihilation in the Sun is in equilibrium. We will focus on the three operators $\bar{\chi} \gamma_\mu \gamma^5 \chi \bar{q} \gamma^\mu \gamma^5 q$, $\bar{\chi} \sigma_{\mu \nu} \chi \bar{q} \sigma^{\mu \nu} q$ and 
$\epsilon^{\mu \nu \rho \sigma} B_\mu (\partial_\nu B_\rho) \bar q \gamma_\sigma \gamma^5 q$, Eqs.~(\ref{eq:fermiwimp}) and (\ref{eq:vectorwimp}), 
which generate spin-dependent interactions in the non-relativistic limit for fermion and vector dark mater, respectively, as demonstrated in Section \ref{sec:eftanalysis}.

First, consider the effective Lagrangian with the axial-vector operator, 
\begin{equation}
\label{eq:lagrangetchannel}
 \mathcal{L}_{\text{eff}, AA} = d_q \bar{\chi} \gamma_\mu \gamma^5 \chi \bar{q} \gamma^\mu \gamma^5 q,
\end{equation}
where $d_q$ has mass-dimension $-2$ and specifies the coupling strength of a quark with flavor $q$. As shown in Section~\ref{sec:BSMWIMP}, the specific interpretation of this effective coupling depends on the underlying UV-complete theory. We will keep $d_q$ as a free parameter to be constrained from the IceCube measurement of the neutrino flux. 

In order to obtain the cross-section for WIMP-nucleon scattering, one has to compute the matrix element 
\begin{equation}\label{eq:matrix_nuc}
 \mathcal{M}_{if, AA} = \sum_{q=u,d,s} d_q \langle N_f , \chi_f | \bar{\chi} \gamma_\mu \gamma^5 \chi \bar{q} \gamma^\mu \gamma^5 q | N_i, \chi_i \rangle,
\end{equation}
where $| N_{f},\chi_{f} \rangle$ and $| N_{i}, \chi_{i} \rangle$  denotes the final and initial  states, respectively. The summation includes the light quarks only, since heavy quarks do not contribute significantly to the nuclear spin. The matrix element in Eq.~(\ref{eq:matrix_nuc}) can be expressed in terms of the spin expectation values of the protons and neutrons in the nucleus~\cite{Jungman1996, Agrawal2010a},
\begin{eqnarray}
 \langle N_f | \bar{q} \gamma^\mu \gamma^5 q | N_i \rangle = && 2 \delta^{\mu j} \Bigl[\langle N_f | (S_p)_j | N_i \rangle \Delta^{(p)}_q \nonumber \\
 &&+ \langle N_f | (S_n)_j | N_i \rangle \Delta^{(n)}_q \Bigr],
\end{eqnarray}
where $(S_{p/n})_j$ denote the spatial components of the proton/neutron spin operators,  
and $\Delta^{(p)}_q$ and $\Delta^{(n)}_q$ are the fractions of the proton or neutron spin carried by quark $q$. 

The expectation values for the proton and neutron spin can be related to the total nuclear angular momentum $J_N$ via
\begin{eqnarray}
 \langle \vec{S_p} \rangle \Delta^{(p)}_q + \langle \vec{S_n} \rangle \Delta^{(n)}_q &=& \Bigl(\frac{\langle S_p \rangle}{J_N} \Delta^{(p)}_q + \frac{\langle S_n \rangle}{J_N} \Delta^{(n)}_q \Bigr) \nonumber \\
 &&\times \langle N_f | \vec{J_N} | N_i \rangle \nonumber \\
 &\equiv& \lambda_q \langle N_f | \vec{J_N} | N_i \rangle,
\end{eqnarray}
where $\langle S_{p/n} \rangle$ are the expectation values of the spin projections onto the z-axis. 

Since the capture of WIMPs in the Sun is dominated by WIMP-proton scattering, these calculations simplify greatly. In this case $J_N = \langle S_{p} \rangle = 1/2$ and $\langle S_{n} \rangle = 0$. For the fractions of the proton spin carried by light quarks, $\Delta^{(p)}_{u,d,s}$, we use the values $\Delta_u = 0.842 \pm 0.012$,  $\Delta_d = -0.427 \pm 0.012$, and $\Delta_s = -0.085 \pm 0.017$, as determined by the HERMES collaboration~\cite{Airapetian2007}.

Squaring the matrix elements, summing and averaging over spins and performing the phase-space integration, one arrives at the WIMP-nucleon scattering cross-section in the non-relativistic limit
\begin{equation}\label{eq:nucxs}
 \sigma_{{\rm SD}, AA} = \frac{4 \mu^2}{\pi} \Bigl[ \sum_{q=u,d,s} d_q \lambda_q \Bigr]^2 J_N(J_N +1),
\end{equation}
where $\mu = m_\chi m_N/(m_\chi + m_N)$ is the reduced mass of the WIMP-nucleus system. Note that the matrix element for a Majorana fermion WIMP contains an additional factor of two compared to the Dirac case, and thus the expression in Eq.~(\ref{eq:nucxs}) should be multiplied by a factor of four when considering Majorana dark matter.

The scattering cross-section for an effective Lagrangian containing the tensor operator, 
\begin{equation}\label{eq:ttlagrangian}
\mathcal{L}_{\text{eff}, TT} = b_q \bar{\chi} \sigma_{\mu \nu} \chi \bar{q} \sigma^{\mu \nu} q,
\end{equation}
is four times that of the axial-vector operator~\cite{Agrawal2010a}. For a Dirac fermion we thus find 
\begin{equation}\label{eq:ttxs}
 \sigma_{{\rm SD}, TT} = \frac{16 \mu^2}{\pi} \Bigl[ \sum_{q=u,d,s} b_q \lambda_q \Bigr]^2 J_N(J_N +1),
\end{equation}
where $b_q$ is a constant with mass dimension $-2$ and describes the strength of the coupling of the tensor interaction for a quark with flavor $q$. Recall that this operator vanishes for a Majorana fermion.

Finally, for a vector boson WIMP, the Lagrangian relevant for spin-dependent interactions is
\begin{equation}
\label{eq:lagrangevecboson}
 \mathcal{L}_{\text{eff}, VB} = c_q \epsilon^{\mu \nu \rho \sigma} \bar q \gamma_\rho \gamma^5 q B_\nu \partial_\sigma B_\mu, 
\end{equation}
resulting in a non-relativistic scattering cross section \cite{Agrawal2010a}
\begin{equation}\label{eq:xsvec}
 \sigma_{{\rm SD}, VB} = \frac{8 \mu^2}{3 \pi} \Bigl[ \sum_{q=u,d,s} c_q \lambda_q \Bigr]^2 J_N(J_N +1). 
\end{equation}
Again, $c_q$ has mass dimension $-2$ and quantifies the strength of the interaction for quark flavor $q$. \newline

In order to be able to compare the IceCube measurements with constraints on the effective theory from the dark matter relic density, we also need to compute the annihilation cross-sections for the operators $\mathcal{O}^{(6)}_{AA}$, $\mathcal{O}^{(6)}_{TT}$ and $\mathcal{O}^{(6)}_{VB}$ described above. These cross sections have been presented in the literature, see \textit{e.g.}\ \cite{Dreiner2013, Zheng2012, Yu2012}, and we will quote the results here for completeness. For the axial-vector operator $\mathcal{O}^{(6)}_{AA}$ in Eq.~(\ref{eq:lagrangetchannel}) the cross-section for the annihilation of Dirac fermion WIMPs into quarks is given by \cite{Zheng2012, Dreiner2013}
\begin{eqnarray}
 \sigma_{{\rm ann}, AA} &=& \frac{1}{4 \pi} \sum_q d^2_q \sqrt{\frac{s-4 m^2_q}{s-4 m^2_\chi}} \nonumber \\
 && \times \Bigl[ s - 4 (m^2_\chi + m^2_q) + 28 \frac{m^2_\chi m^2_q}{s} \Bigr].
\end{eqnarray}
Similarly, the annihilation cross-section for the tensor operator $\mathcal{O}^{(6)}_{TT}$ in Eq.~(\ref{eq:ttlagrangian}) is given by \cite{Zheng2012}
\begin{eqnarray}
 \sigma_{{\rm ann}, TT} &=& \frac{1}{2 \pi} \sum_q b^2_q \sqrt{\frac{s-4 m^2_q}{s-4 m^2_\chi}} \nonumber \\
 && \times \Bigl[ s + 2 (m^2_\chi + m^2_q) + 40 \frac{m^2_\chi m^2_q}{s} \Bigr].
\end{eqnarray}
Finally, for the case of a real vector boson WIMP, the annihilation cross-section for the operator $\mathcal{O}^{(6)}_{VB}$ in Eq.~(\ref{eq:lagrangevecboson}) reads \cite{Yu2012}
\begin{eqnarray}
 \sigma_{{\rm ann}, VB} &=& \frac{1}{9 \pi m^2_\chi} \sum_q c^2_q \sqrt{(s-4 m^2_q)(s-4 m^2_\chi)} \nonumber \\
 && \times \Bigl[ s - 4 (m^2_\chi + m^2_q) + 28 \frac{m^2_\chi m^2_q}{s} \Bigr].
\end{eqnarray}
In all three terms, the sum runs over all annihilation channels that are kinematically accessible, \textit{i.e.}\ those where $m_\chi \geq m_q$.

The expected present day dark matter abundance can be calculated using the standard approach described in \cite{Bertone2005, Jungman1996} and is given by
\begin{equation}\label{eq:DMrelic}
 \Omega_\chi \text{h}^2 = \frac{1.07 \times 10^{9} \; \text{GeV}^{-1}}{M_{\rm Pl}} \frac{x_F}{\sqrt{g_\star}} \frac{1}{a + 3b/x_F},
\end{equation}
where $M_{\rm Pl} = 1.22 \times 10^{19} \; \text{GeV}$ is the Planck-mass, $g_\star$ is the number of relativistic degrees of freedom at the time of WIMP-decoupling, $x_F = m_\chi/T_F$ is the inverse freeze-out temperature scaled by the WIMP mass, and $a$ and $b$ are given by the coefficients of the low-velocity expansion of the annihilation cross-section $\sigma_{\rm ann} v \simeq a + b v^2$. In most models the value for the inverse freeze-out temperature hardly varies and typically lies between $20 < x_F < 30$ \cite{Zhang2011, Jungman1996}. The corresponding number of degrees of freedom is given in Ref.~\cite{Coleman2003} and is typically somewhere between $80$ and $100$ for several orders of magnitude in temperature. In order to find the coefficients $a$ and $b$ one has to perform the non-relativistic approximation $s \simeq 4m^2_\chi + m^2_\chi v^2 + 3/4 m^2_\chi v^4$. The corresponding expressions have been calculated in \cite{Zheng2012,Yu2012,Dreiner2013} and read
\begin{eqnarray}
\label{eq:sigmav}
 \sigma_{{\rm ann}, AA} v &\simeq& \frac{3 m^2_\chi}{2 \pi} \sum_q d^2_q \nonumber \\ 
  && \times \Bigl[\sqrt{1 - \xi^2} \xi^2 + \frac{8 - 28 \xi^2 + 23 \xi^4}{24 \sqrt{1 - \xi^2}} v^2 \Bigr] \nonumber \\
  \vspace{15pt}
 \sigma_{{\rm ann}, TT} v &\simeq& \frac{6 m^2_\chi}{\pi} \sum_q b^2_q \sqrt{1 - \xi^2} (1 + 2\xi^2) \nonumber \\ 
 && \times \Bigl[1 + \frac{-2 - 17 \xi^2 + 28 \xi^4}{24 (1 - \xi^2)(1+2\xi^2)} v^2 \Bigr] \nonumber \\
 \vspace{15pt}
 \sigma_{{\rm ann},VB} v &\simeq& \frac{2 m^2_\chi}{3 \pi} \sum_q c^2_q \sqrt{1 - \xi^2} \xi^2 v^2, 
\end{eqnarray}
where we have defined $\xi = m_q/m_\chi$.


\section{Results}
\label{sec:results}

\begin{figure*}
\includegraphics[width=0.49\textwidth]{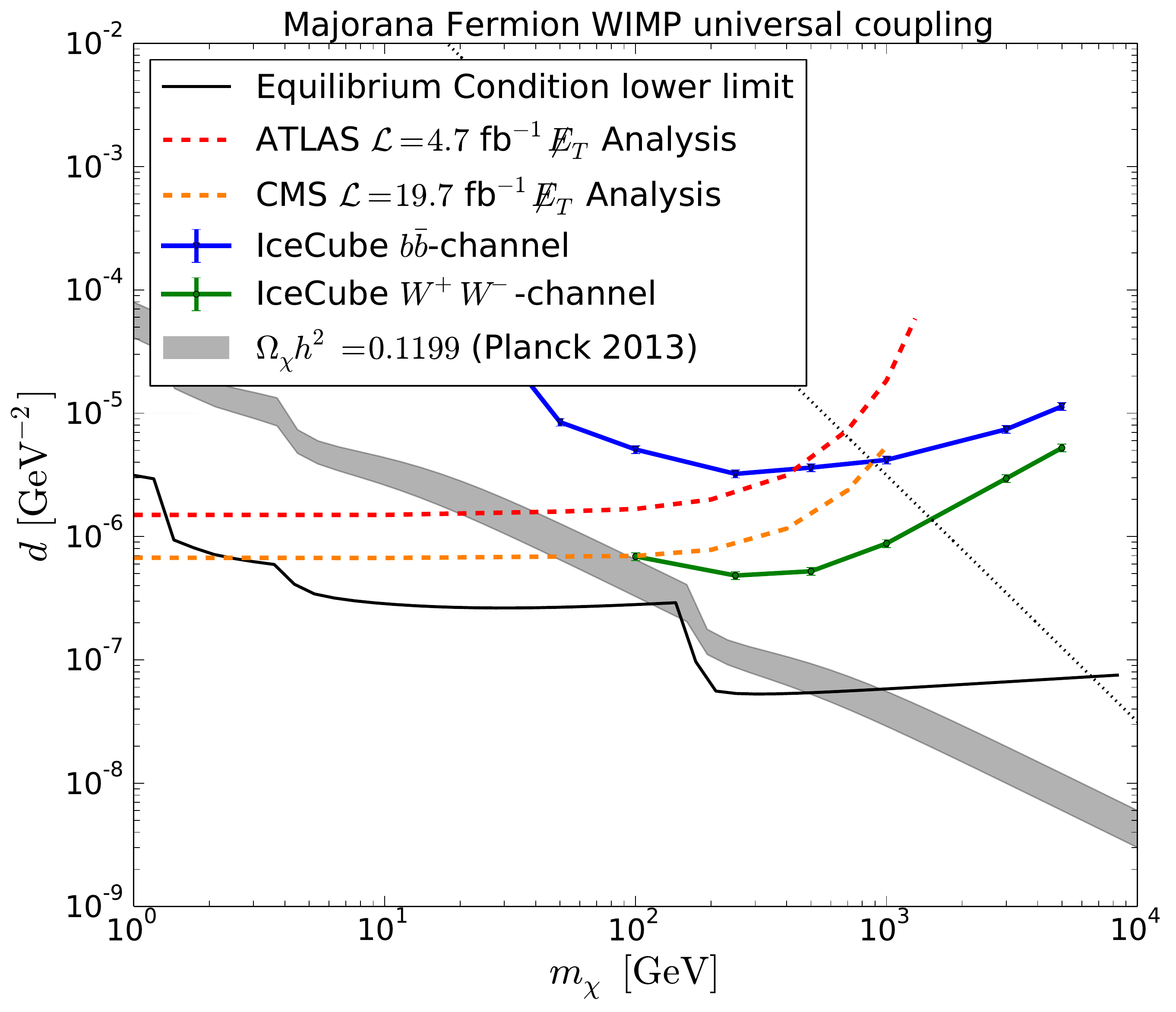}
\includegraphics[width=0.49\textwidth]{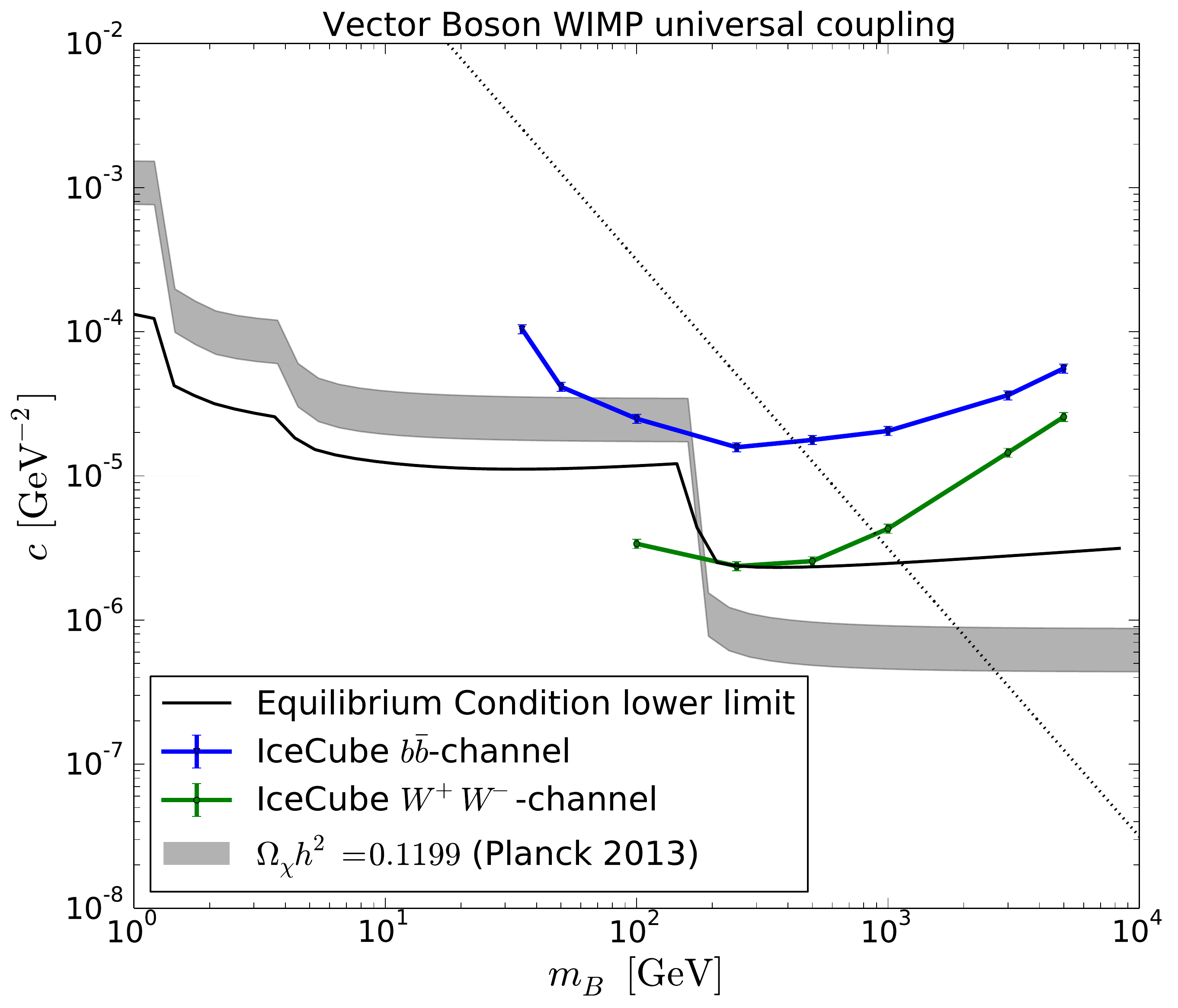}
\includegraphics[width=0.49\textwidth]{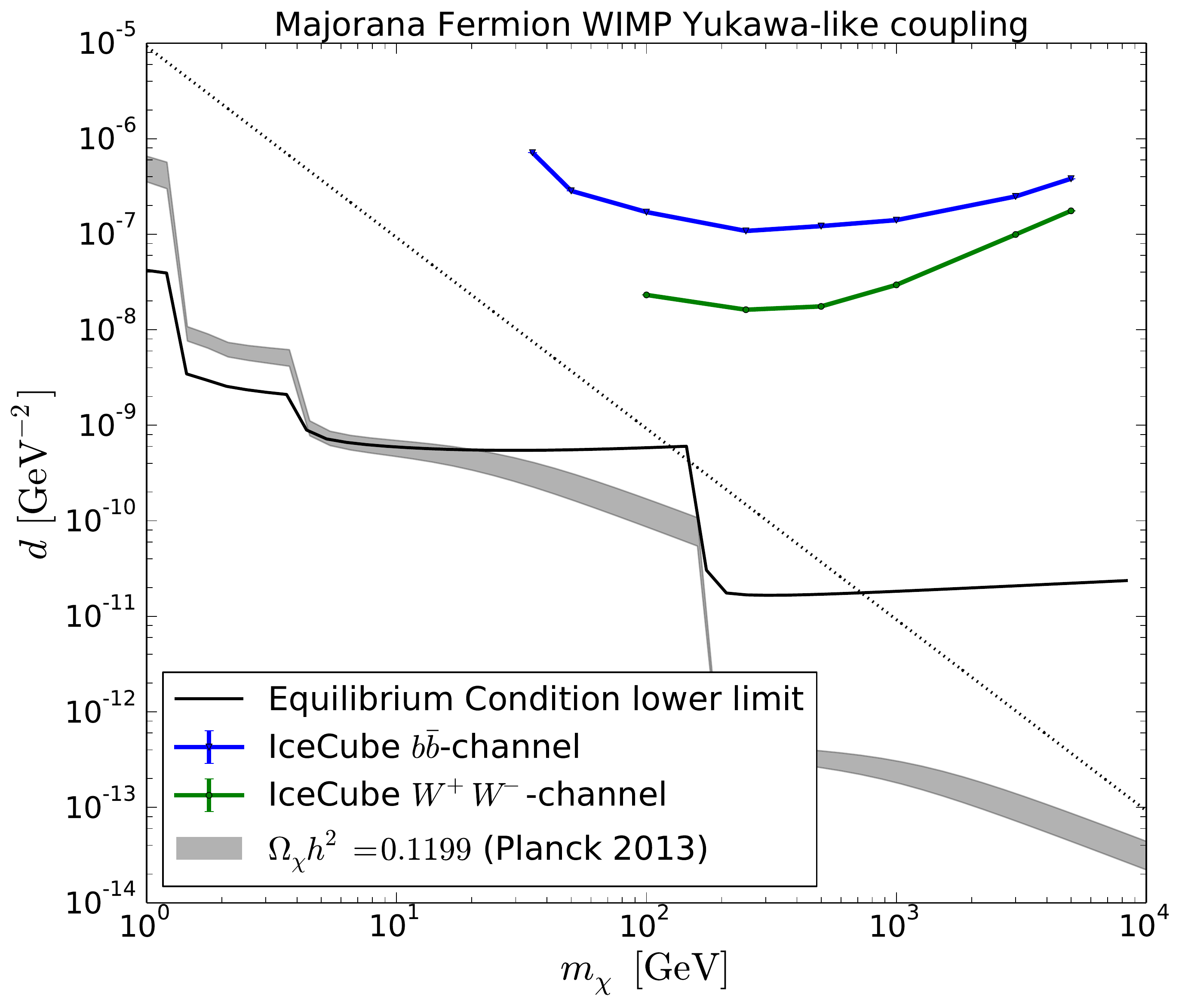}
\includegraphics[width=0.49\textwidth]{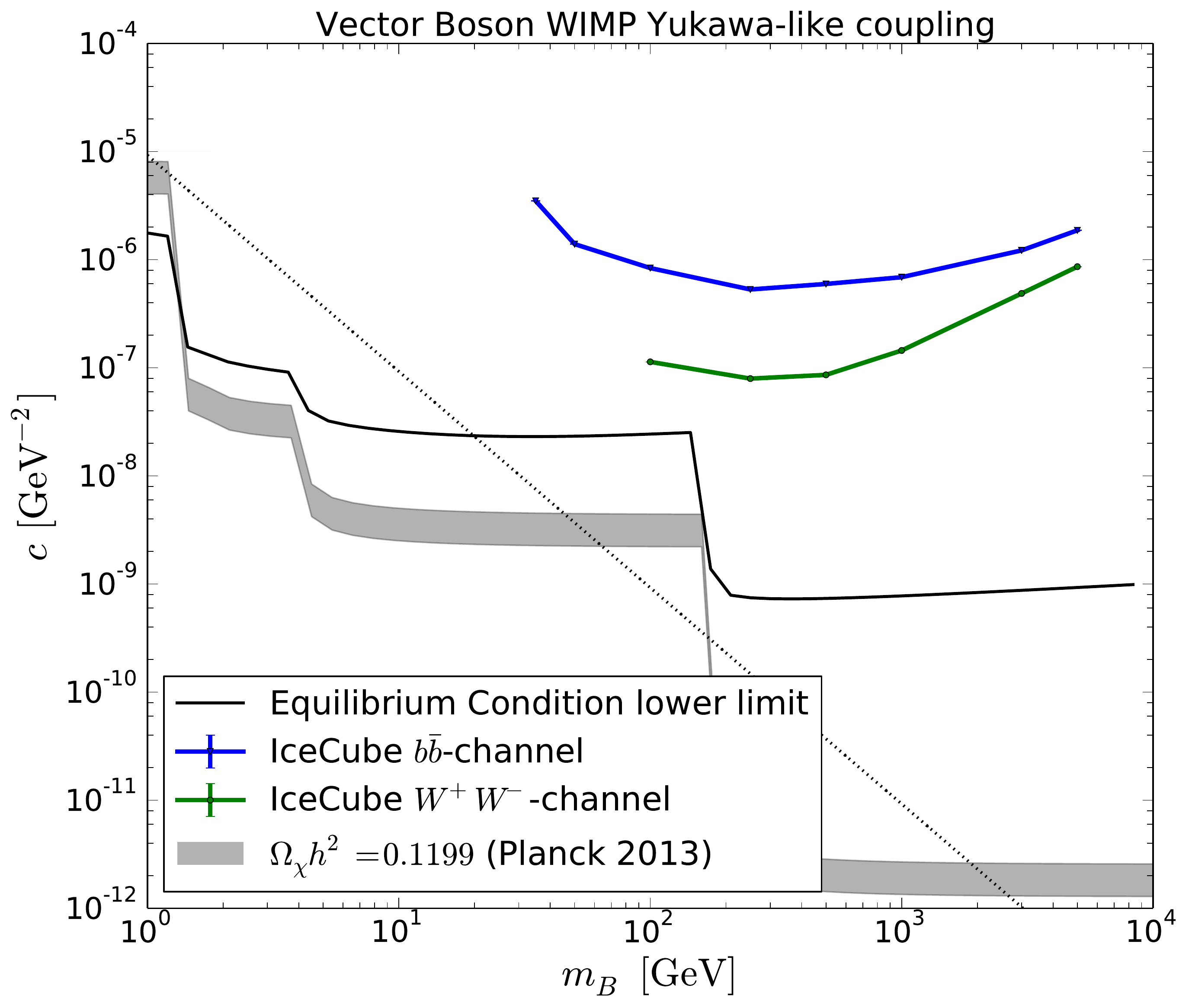}
\caption[Upper limits on the effective coupling constants]{(Top left): Comparison of upper limits on the universal effective coupling constant $d$ of the axial-vector operator for a Majorana fermion WIMP. The blue (green) lines show the upper limits on the effective coupling from spin-dependent scattering cross-sections given by the IceCube search for dark matter annihilation in the Sun for soft or hard neutrino spectra corresponding to the $b \bar b$-channel ($W^{+}W^{-}$-channel) \cite{Aartsen2013}. The red (orange) dashed lines show the 90\% CL upper limits from dark matter searches in mono-jet events at the LHC by the ATLAS (CMS) collaborations~\cite{Aad2013, CMS2014}. The gray band corresponds to the value of the coupling needed in order to account for the present day relic abundance of dark matter as measured by the Planck satellite \cite{Ade2013}. The black line indicates the upper limit for equilibrium between capture and annihilation in the Sun as derived from Eq.~(\ref{eq:equil}). The region above the dotted line corresponds to values of the parameters for which the effective field theory approach to dark matter annihilation is unreliable (see discussion in \ref{sec:results}). (Top right, bottom left, bottom right): Upper limits on the effective couplings for a vector boson WIMP with universal couplings, as well as for Majorana fermion and vector boson WIMPs with Yukawa-like coupling, respectively. Note that in these cases there are no corresponding upper limits given by LHC searches.}
\label{fig:couplinglimits}
\end{figure*}

In order to be able to derive upper limits on the effective coupling constants $d_q$, $b_q$ and $c_q$ in the effective Lagrangians, Eqs.~(\ref{eq:lagrangetchannel}), (\ref{eq:ttlagrangian}) and (\ref{eq:lagrangevecboson}), and the corresponding expressions for the scattering cross-sections, Eqs.~(\ref{eq:nucxs}), (\ref{eq:ttxs}) and (\ref{eq:xsvec}), respectively, we have to make an assumption about the coupling strength of dark matter to individual quark flavors. We consider two benchmark scenarios: one with universal couplings, where we assume that the coupling strength of the effective operators is the same for all quarks, $d_q=d$, $b_q = b$ and $c_q = c$. The second benchmark scenario assumes Yukawa-like couplings, with coupling strengths which scale with the mass of the quark $d_q = d \; m_q/m_e$, and similarly for $b_q$ and $c_q$, where $m_e$ is the mass of the electron and $d$, $b$ and $c$  are universal factors. This second coupling scenario could be indicative of an interaction mediated by some kind of Higgs particle, even though it is not obvious how that might be realised in a UV-complete model for the cases we are considering here.  We will remain agnostic about the origin of the Yukawa-like couplings and simply consider it as a benchmark scenario. The resulting limits on the effective couplings for both the universal and the Yukawa-like coupling type are shown in FIG.~\ref{fig:couplinglimits}.

For the fermionic WIMP we have assumed $\chi$ to be a Majorana fermion, since in our simplified models this naturally leads to predominantly spin-dependent interactions. Also, results for Majorana WIMPs can be interpreted in the context of full-fledged new physics models like the minimal supersymmetric model \cite{Ellis:1983ew}, in which dark matter is in general the lightest neutralino. We have neglected the tensor operator in this analysis, since it vanishes for Majorana fermions. For the vector boson WIMP, we also focus on models which are able to naturally create predominantly spin-dependent interactions and therefore consider the case of a real vector particle. 

The top left panel in FIG.~\ref{fig:couplinglimits} shows the IceCube upper limits on the effective coupling $d$ for Majorana dark matter with universal couplings to quarks. We provide a weaker  limit based on a soft neutrino spectrum as expected from WIMP annihilation into bottom quarks, and a stronger limit based on a harder neutrino spectrum corresponding to annihilation into $W$ bosons or $\tau$ leptons, see Section~\ref{sec:icresults}. As we assume a minimal dark matter model where the WIMP couples to quarks only, it is not obvious how one could generate a neutrino spectrum as hard as that for annihilation into $W$ bosons. Note, however, that final states with additional $W$ and $Z$ bosons, $\chi\chi\to q\bar{q}(\bar{q}')+ Z(W)$,  are strongly enhanced for Majorana dark matter annihilation into light fermions, as the emission of vector bosons lifts the helicity suppression $\propto m_q^2/m_\chi^2$ of the  $\chi\chi\to q\bar{q}$ cross section~\cite{Bringmann2014, Garny2011, Bell2011}. While the neutrino spectrum from the decay of secondary $W$ and $Z$ bosons differs from that of primary $W$ bosons \cite{Bell2011a}, we can still expect a spectrum harder than that resulting from primary bottom quarks. We will consider the hard spectrum from annihilation into $W^{+}W^{-}$ pairs as the most optimistic scenario for our analyses, expecting the actual neutrino spectrum to lie somewhere in between that of the $b \bar b$- and $W^{+}W^{-}$-channels. Note that electroweak radiation is, in general, not enhanced by helicity factors for the case of vector WIMP annihilation, so that a soft neutrino spectrum may be more appropriate for these scenarios.

We compare our limits from dark matter annihilation in the Sun with constraints from WIMP searches at the LHC as reported by ATLAS \cite{Aad2013} and CMS \cite{CMS2014}, and with constraints set by the dark matter relic density derived from Eq.~(\ref{eq:DMrelic}). Before we comment on the complementarity of the various searches and constraints, let us first discuss the validity of the effective field theory approach for the different types of calculations. The elastic WIMP-nucleus cross sections which determine the WIMP capture rate in the Sun are evaluated at zero-momentum transfer, and thus the effective field theory approach is valid for any relevant size of effective coupling or dark matter mass. To set limits in the parameter space of WIMP mass and effective coupling from the IceCube measurement of the neutrino flux, we have to assume that WIMP capture and annihilation are in equilibrium. To test this assumption we have evaluated the equilibrium condition $\sqrt{C_{\astrosun} C_A} \, t \gg 1$ by inserting our results for the elastic scattering and annihilation cross sections into Eq.~(\ref{eq:equil}). The result is presented in FIG.~\ref{fig:couplinglimits} as an equilibrium condition lower limit, \textit{i.e.}\ the equilibrium condition holds for effective couplings above the line. Our  upper limits from the IceCube analysis are well above the equilibrium condition lower limit, which a posteriori justifies our assumption of equilibrium when relating limits on the annihilation rate and limits on scattering cross-sections. Note that in general also the parameter space below the equilibrium limit can be probed, however a more sophisticated analysis including the $\text{tanh}^2$-suppression of the annihilation rate would be necessary in such cases, see 
Eq.~(\ref{eq:tanhsq}).

Our estimate of the equilibrium condition lower limit rests on the validity of the effective field theory approach for the calculation of the annihilation cross section $\langle \sigma_{\rm ann} v\rangle$. Recall that we need to require $E \lesssim \Lambda$ for the expansion in inverse powers of $\Lambda$ to be reliable. With $d = f/\Lambda^2$, and $E = 2 m_\chi$ for $s$-channel WIMP annihilation, this leads to $4 m_\chi^2 \lesssim f/d$. For a perturbative coupling we require $g^2 \sim f \lesssim 4\pi$ and thus we arrive at the condition
\begin{equation}
\label{eq:eftcondition1}
 d \lesssim \frac{\pi}{m^2_\chi}.
\end{equation}
For a coupling $d$ above $\pi/m_\chi^2$, our effective field theory calculation of the WIMP annihilation cross section is not reliable and we cannot properly evaluate the equilibrium condition lower limit. The region $d \ge \pi/m^2_\chi$ is marked by the dotted line in FIG.~\ref{fig:couplinglimits}. We find that our analysis is self-consistent and robust up to dark matter masses of $m_\chi \lesssim 1$\,TeV. However, we can expect our limits to hold for larger WIMP masses in a wide class of UV-complete models. Consider, for example, the simplified model with a vector mediator specified  by the Lagrangian in Eq.~(\ref{eq:lagrangian_FV}), which generates the axial-vector coupling that we consider here. In such models, WIMP annihilation proceeds through the $s$-channel exchange of the mediator. The breakdown of the effective field theory description corresponds to annihilation energies near or above the mediator mass, where the on-shell production of the mediator particle becomes possible. However, near such resonances, the simplified model would lead to an annihilation cross section which is enhanced compared to that of the effective field theory, at least as long as the width of the mediator is smaller than its mass. A larger annihilation cross section, however, implies a larger factor $\sqrt{C_{\astrosun} C_A} \, t $, such that the equilibrium condition lower limit calculated using the effective field theory is a conservative estimate. A more quantitative analysis requires a model-specific calculation, see \textit{e.g.}~Ref.~\cite{Ellis:2009ka} for a study within the minimal supersymmetric model. 

The gray band in FIG.~\ref{fig:couplinglimits} corresponds to the value of the effective coupling which would yield the right dark matter abundance of $\Omega_\chi \text{h}^2 = 0.1199$ as measured by the Planck satellite \cite{Ade2013}. The width of the band arises from varying the inverse freeze-out temperature $x_F$ between 20 and 30. This curve serves as an lower bound on the effective coupling. For couplings above this curve, the dark matter abundance would be lower than the measured value, which is acceptable if we allow for several WIMPs or other sources of dark matter, like axions. For values below the curve, the WIMP would provide more than the observed abundance, and the corresponding model is generally considered excluded in a scenario where there are no additional mechanisms, such as co-annihilation of nearly mass-degenerate WIMPs \cite{Jungman1996}. For Majorana fermion WIMPs the upper limits placed by our IceCube analysis are above the gray band and thus consistent with the dark matter relic density. 

In FIG.~\ref{fig:couplinglimits} we also show limits from collider searches for dark matter as reported by the ATLAS and CMS collaborations \cite{Aad2013, CMS2014}. These limits are competitive with the IceCube limits for WIMP masses $m_\chi \lesssim 850 \; \text{GeV}$, even if we conservatively assume a soft neutrino spectrum. Note, however, that the validity of the effective field theory approach for dark matter searches at the LHC has been questioned recently by a number of authors, see \textit{e.g.}\,\cite{Buchmueller2013, Busoni2014, Shoemaker2012}, and we will comment on the implications later in this section. 

In the lower left panel of FIG.~\ref{fig:couplinglimits} we show the corresponding analysis for the Yukawa-like benchmark scenario with $d_q = d \; m_q/m_e$. The limits are about two orders of magnitude weaker than those for universal couplings. For Yukawa-type models the validity of the effective field theory calculation of the annihilation cross is 
\begin{equation}
\label{eq:eftcondition2}
 d \lesssim \frac{m_e \pi}{m_q m^2_\chi},
\end{equation}
where we have set $m_q = m_t = 173.5 \; \text{GeV}$ to account for the most conservative scenario. The region above the dotted line corresponds to couplings $d \ge m_e \pi / (m_t m^2_\chi)$ where the effective field theory analysis of WIMP annihilation is doubtful. Thus, if equilibrium of dark matter annihilation and capture in the Sun holds, we can probe a parameter space of the effective theory with the IceCube analysis, which is not accessible to corresponding indirect detection searches. As we have argued above, our equilibrium condition lower limit should be conservative in a wide class of UV-complete models, and the limits presented in FIG.~\ref{fig:couplinglimits} should, accordingly, be sound. 

Very similar conclusions hold for the case of vector dark matter presented in the right panels of FIG.~\ref{fig:couplinglimits}. 

We can convert the upper limits on the effective coupling $d$ presented in FIG.~\ref{fig:couplinglimits} into lower limits on the mass of the mediator particle in the context of the simplified models constructed in Section~\ref{sec:BSMWIMP}. Comparing the effective Lagrangian in Eq.~(\ref{eq:lagrangetchannel}) and the second term in Eq.~(\ref{eq:axialtchannel}) we deduce that 
\begin{equation}
 d = \frac{g_A g'_A }{m^2_V}.
\end{equation}
If we furthermore require perturbative couplings $g_A, g'_A < \sqrt{4 \pi}$, and assume for the sake of simplicity that the mediating vector boson couples with the same strength to the WIMP and to quarks, $g_A = g'_A$, we arrive at the relation
\begin{equation}
 m_V > \sqrt{\frac{4 \pi}{d}}.
\end{equation}

\begin{figure}
\includegraphics[width=0.5\textwidth]{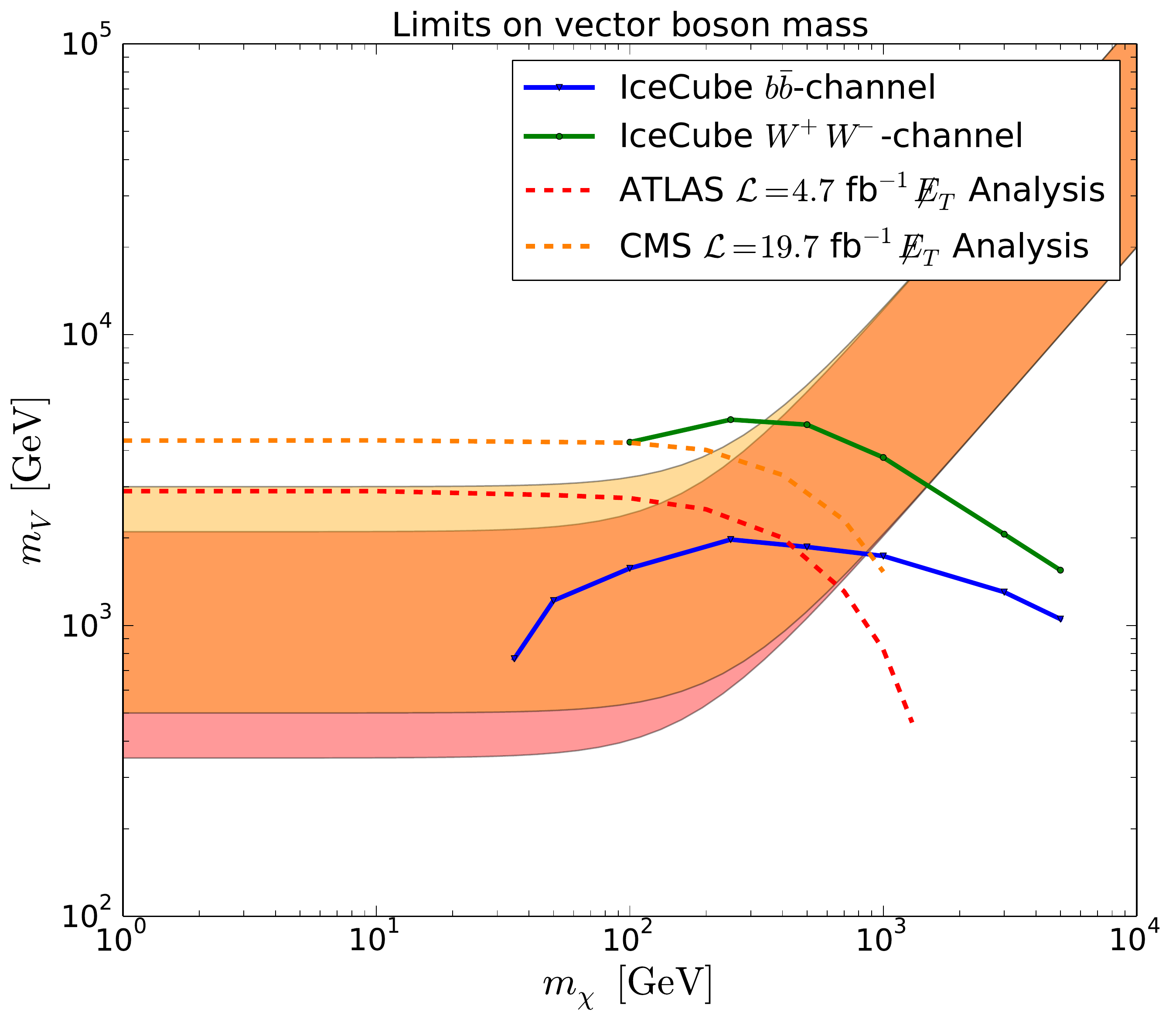}
\caption[Lower bounds on the mass of a heavy vector boson mediator]{Lower bounds on the mass of the mediating vector boson $m_V$ for Majorana dark matter with universal axial-vector couplings to quarks. The blue (green) curve corresponds to limits set by the IceCube solar WIMP analysis for soft (hard) neutrino spectra corresponding to the $b \bar b$-channel ($W^{+}W^{-}$-channel). 
The red (orange) dashed lines show lower limits from dark matter searches in mono-jet events at the LHC by the ATLAS (CMS) collaborations~\cite{Aad2013, CMS2014}. The red (orange) shaded area corresponds to the region given in \ref{eq:resonance}, where the effective field theory approach significantly underestimates the actual LHC limits provided by ATLAS (CMS), as described in Ref.~\cite{Buchmueller2013}.}
\label{fig:vectormass}
\end{figure}

The resulting lower bounds on the mediator mass are shown in FIG.~\ref{fig:vectormass} for the case of universal couplings. For a WIMP mass of $m_\chi \approx 1$\,TeV, the IceCube data can exclude vector boson mediators lighter than approximately $2$ and $5 \; \text{TeV}$, for a soft or hard neutrino spectrum, respectively.  Assuming that the couplings $g_A, g'_A $ can be as large as $4 \pi$ corresponds to the most optimal scenario, and the limits would be weaker if the couplings of the underlying theory would be smaller. However, the LHC bounds on the mediator mass would scale in the same way so that the relative sensitivity of IceCube and LHC searches does not change. 

The ATLAS and CMS analyses of mono-jet events assume that the creation of WIMPs is described by one of several effective operators. For each one of these operators a signal prediction is made and then compared with the background expectation in several signal regions, mainly characterised by the missing energy in the final state and the transverse momentum of a leading hadronic jet \cite{Aad2013, CMS2014}.

For the axial-vector operator the final analysis compares the number of measured events and events expected from background in a signal region which requires $p^{\text{jet}}_T, E^{\text{miss}}_T > 350 \; \text{GeV}$ for the ATLAS search and $E^{\text{miss}}_T > 500 \; \text{GeV}$ and $p^{\text{jet}}_T > 110 \; \text{GeV}$ for the CMS search. Since no significant excess over the background expectation was observed, 90\% CL lower limits on the suppression scale $\Lambda$ of the effective operator can be derived from this analysis. Note that the LHC collaborations report limits on the axial-vector coupling for Dirac dark matter. We have thus rescaled the numbers provided in Refs.~\cite{Aad2013, CMS2014} by a factor of 2 to account for the larger cross section for Majorana dark matter production. 

It is important to note, however, that because of the potentially large momentum transfer in high-energy collisions, the mass of the mediating particle should be in the TeV-range for the effective field theory to be reliable at the LHC~\cite{Busoni2014, Buchmueller2013}. Ref.~\cite{Buchmueller2013} compares the LHC limits on the suppression scale $\Lambda$ set by the effective field theory approach with limits from a simplified model, for different ranges of the mediator mass. For very large mediator masses, $m_\text{med} \gtrsim 2.5 \; \text{TeV}$, the effective field theory limits reproduce the limits from the simplified model quite well. For lighter mediators, the effective field theory either underestimates the actual limit on $\Lambda$ when the cross section is resonantly enhanced, or it overestimates the limit because the predicted missing energy distribution is too soft compared to the simplified model. Ref.~\cite{Buchmueller2013} provides ``rule of thumb'' approximations for the values of mediator masses at which the transition to the resonant enhancement region occurs. This region depends on the cut applied on the missing transverse energy in the final state and is given by
\begin{equation}
\label{eq:resonance}
\sqrt{4 m^2_\chi + \slashed{E}^2_T} \lesssim m_\text{med} \lesssim 6 \sqrt{4 m^2_\chi + \slashed{E}^2_T}. 
\end{equation}
Above the upper bound, \textit{i.e.}\ for heavy mediators, the effective field theory approach is reliable, below the lower bound, the effective field theory limits overestimate the simplified model limits. Within the bounds, the cross section is resonantly enhanced, so that the effective field theory limits underestimate the limits within the simplified model. In FIG.~\ref{fig:vectormass} the region of resonant enhancement for $\slashed{E}_T = 350 \; \text{GeV}$ and $\slashed{E}_T = 500 \; \text{GeV}$ are indicated by the red and orange bands, respectively. In this region we would thus expect the limits set by LHC searches to be stronger than the ones presented in the figure. However, the limits provided by IceCube extend to WIMP masses beyond the current LHC sensitivity, and beyond the region of resonant cross section enhancement, and thus they are comparable in sensitivity and complimentary in WIMP mass range to the LHC results. 


\section{Astrophysical uncertainties}
\label{sec:uncertainties}

The two largest astrophysical uncertainties in the calculation of WIMP capture rate are the local WIMP density and the WIMP velocity distribution. As was shown in Section~\ref{sec:intro}, assuming equilibrium the total capture rate for WIMPs in the Sun is given by 
\begin{equation}
 C_{\astrosun} = 2 \Gamma_A \propto \sigma_{\rm SI/SD} \int\! {\rm d}u \, f(u).
\end{equation}
The normalisation of the velocity distribution of WIMPs in the halo, $f(u)$, depends on the local dark matter density $f(u) \propto \rho_{\rm loc}$. For any measured value of the annihilation rate in the Sun $\Gamma_A$, the derived limit on the scattering cross-section $\sigma_{\rm SI/SD}$ will be $\propto 1/\int \! {\rm d}u \,f(u) \propto 1/\rho_{\rm loc}$. Since the effective coupling depends on $\sqrt{\sigma_{\rm SI/SD}}$, the impact of the local dark matter density is not very significant. The original IceCube analysis of solar WIMP annihilation \cite{Aartsen2013} had assumed the canonical value of $\rho_{\rm loc} = 0.3 \; \text{GeV} \, \text{cm}^{-3}$. Choosing an extreme scenario of $\rho_{\rm loc} = 0.9 \; \text{GeV} \, \text{cm}^{-3}$ \cite{Garbari2012} would only lead to a $\sim 40 \%$ improvement in the limit on the effective coupling.

\begin{figure}
 \includegraphics[width=0.5\textwidth]{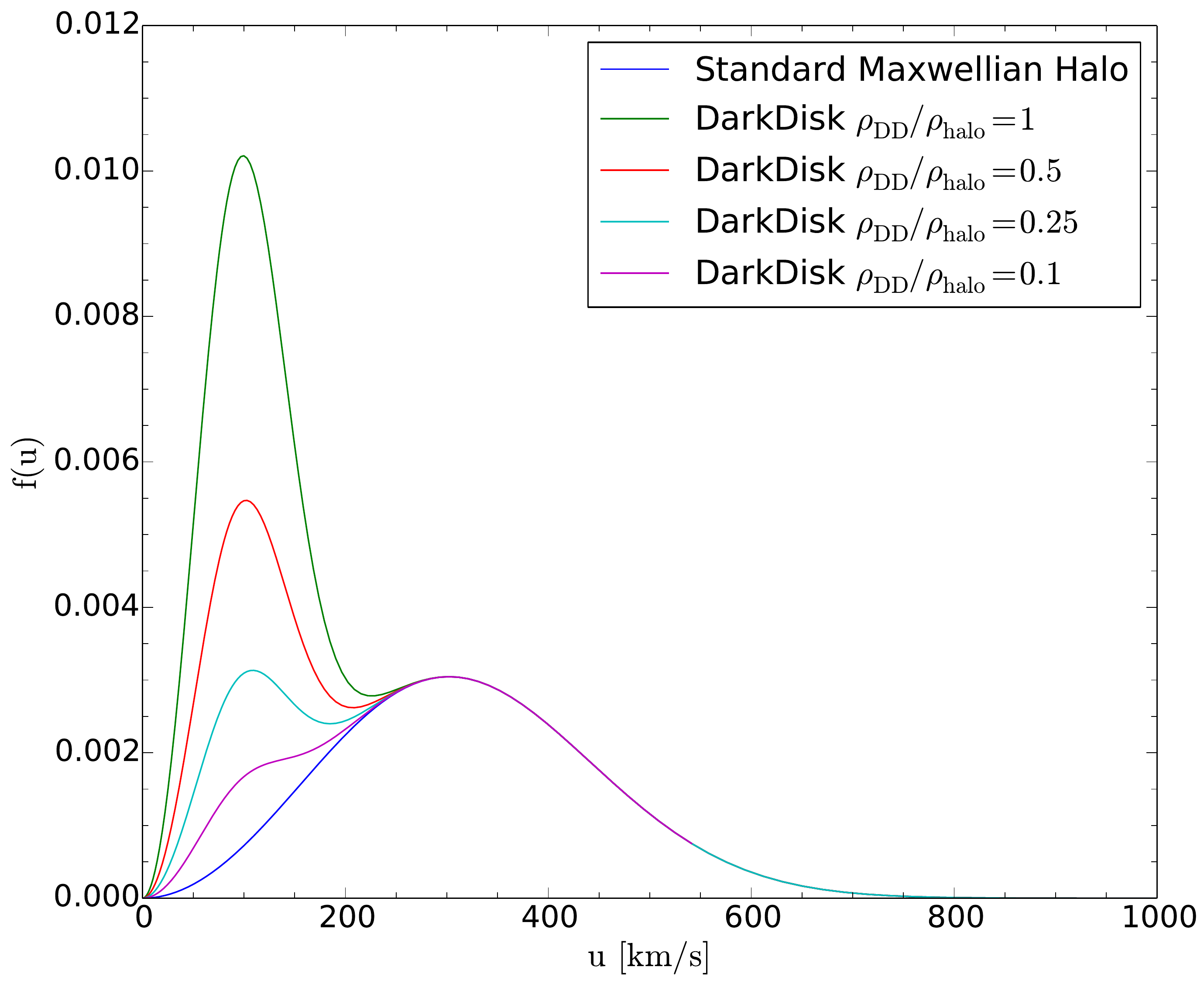}
 \caption[Velocity distributions of WIMPs in the halo]{Comparison of WIMP velocity distributions in the galactic halo for different scenarios. The standard Maxwellian Halo corresponds to a Maxwell-Boltzmann distribution with a velocity dispersion of $\bar v = 270 \; \text{km} \, \text{s}^{-1}$ transformed into the rest frame of the Sun moving through the halo with $v_{\astrosun} = 220 \; \text{km} \, \text{s}^{-1}$. Velocity distributions of dark matter disc scenarios assume a second WIMP population with a relative velocity of $70 \; \text{km} \, \text{s}^{-1}$ and are shown for the extreme cases $\rho_{\rm disc}/\rho_{\rm halo} = 0.1$ and $\rho_{\rm disc}/\rho_{\rm halo} = 1$, as well as intermediate values of $\rho_{\rm disc}/\rho_{\rm halo} = 0.25$ and $\rho_{\rm disc}/\rho_{\rm halo} = 0.5$.}
 \label{fig:veldist}
\end{figure}

The effect of the WIMP velocity distribution itself however can be quite significant. Due to the integration over WIMP velocities, the capture rate is very sensitive to the low-velocity part of $f(u)$ since the capture probability for low velocity WIMPs is significantly higher. Often, the velocity distribution has been modelled as a Maxwell-Boltzmann distribution with a dispersion of $\bar v = 270 \; \text{km} \, \text{s}^{-1}$~\cite{Choi2013}. However, recent simulations of the galactic halo have produced velocity distributions that deviate from the Standard Maxwellian Halo (SMH) \cite{Mao2013, Ling2009, Vogelsberger2009}. It was shown in Ref.~\cite{Choi2013} that these velocity distributions have an impact on the capture rate at a level of up to $\sim 20 \%$. Another interesting scenario presented in \cite{Choi2013} is the case of a co-rotating dark matter disc in the plane of our galaxy. As has been shown by recent N-body simulations which include the effects of baryons on the WIMP distribution in the halo, a co-rotating dark matter disc could form with density fractions ranging from $\rho_{\rm disc}/\rho_{\rm halo} = 0.1$ \cite{Pillepich2013} to $\rho_{\rm disc}/\rho_{\rm halo} = 1$ \cite{Read2009}. The resulting velocity distributions for a dark matter disc co-rotating with a relative speed of $70 \; \text{km} \, \text{s}^{-1}$ are shown in FIG.~\ref{fig:veldist}. 

\begin{figure*}
 \includegraphics[width=0.49\textwidth]{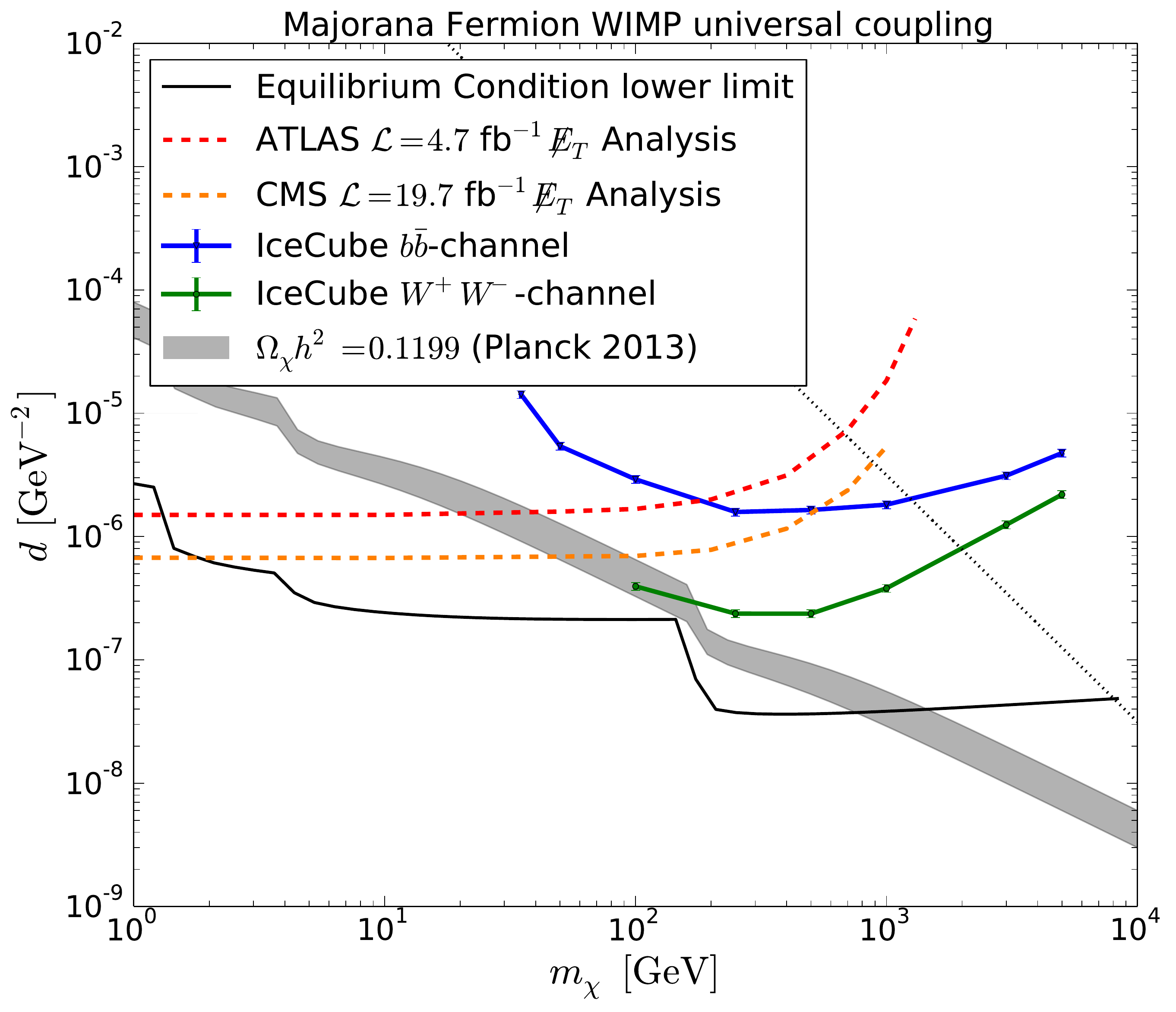}
 \includegraphics[width=0.49\textwidth]{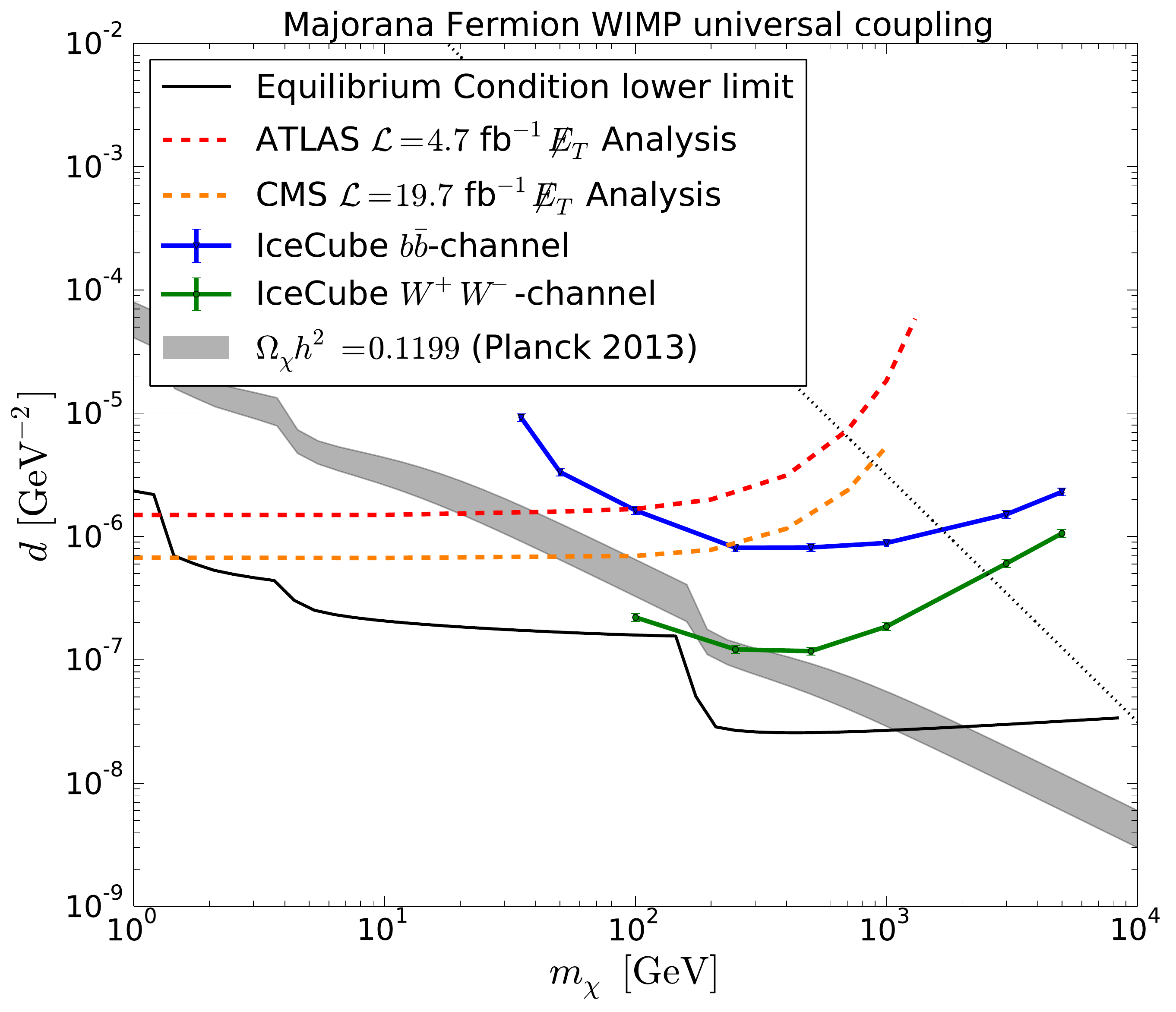}
 \includegraphics[width=0.49\textwidth]{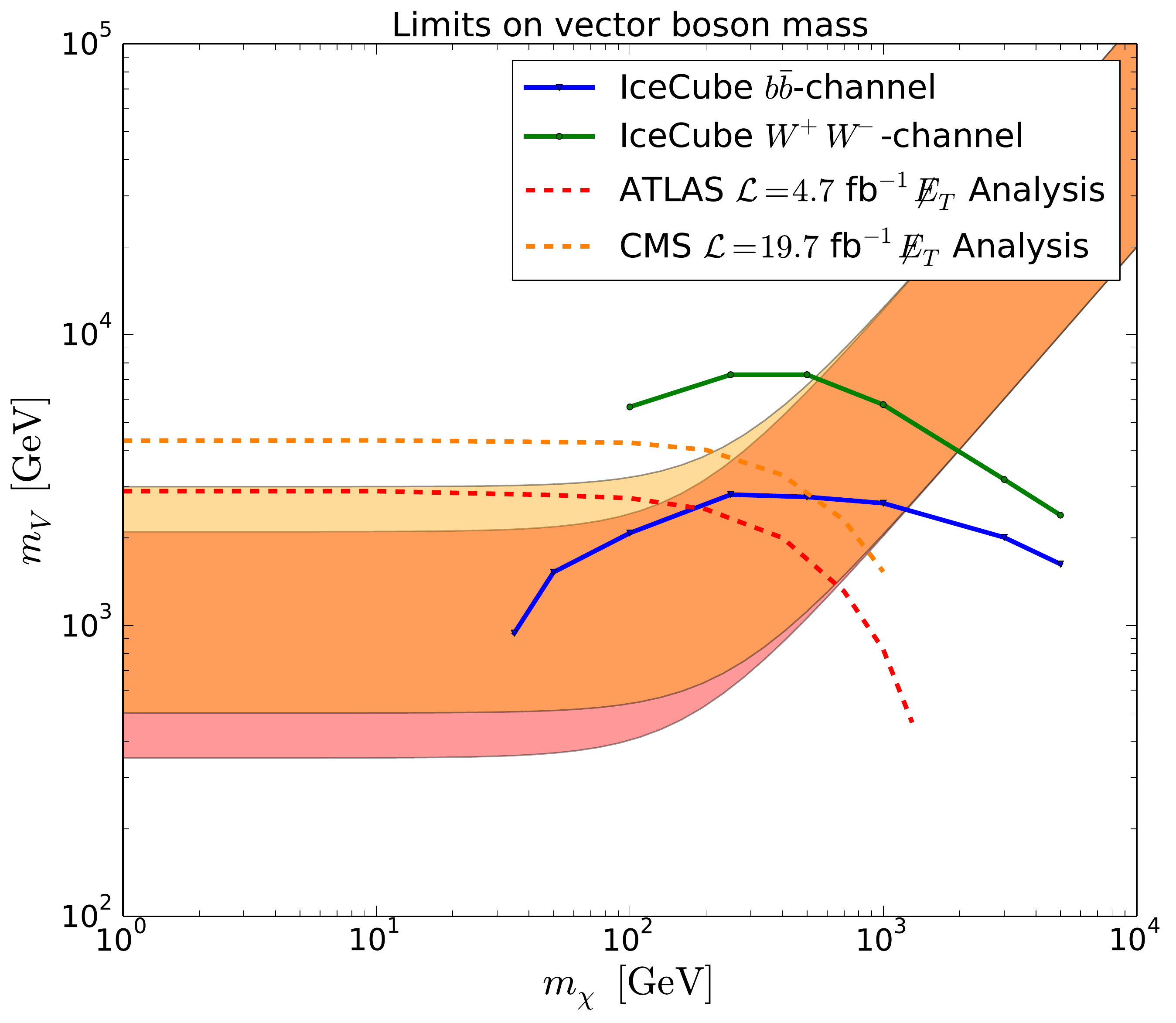}
 \includegraphics[width=0.49\textwidth]{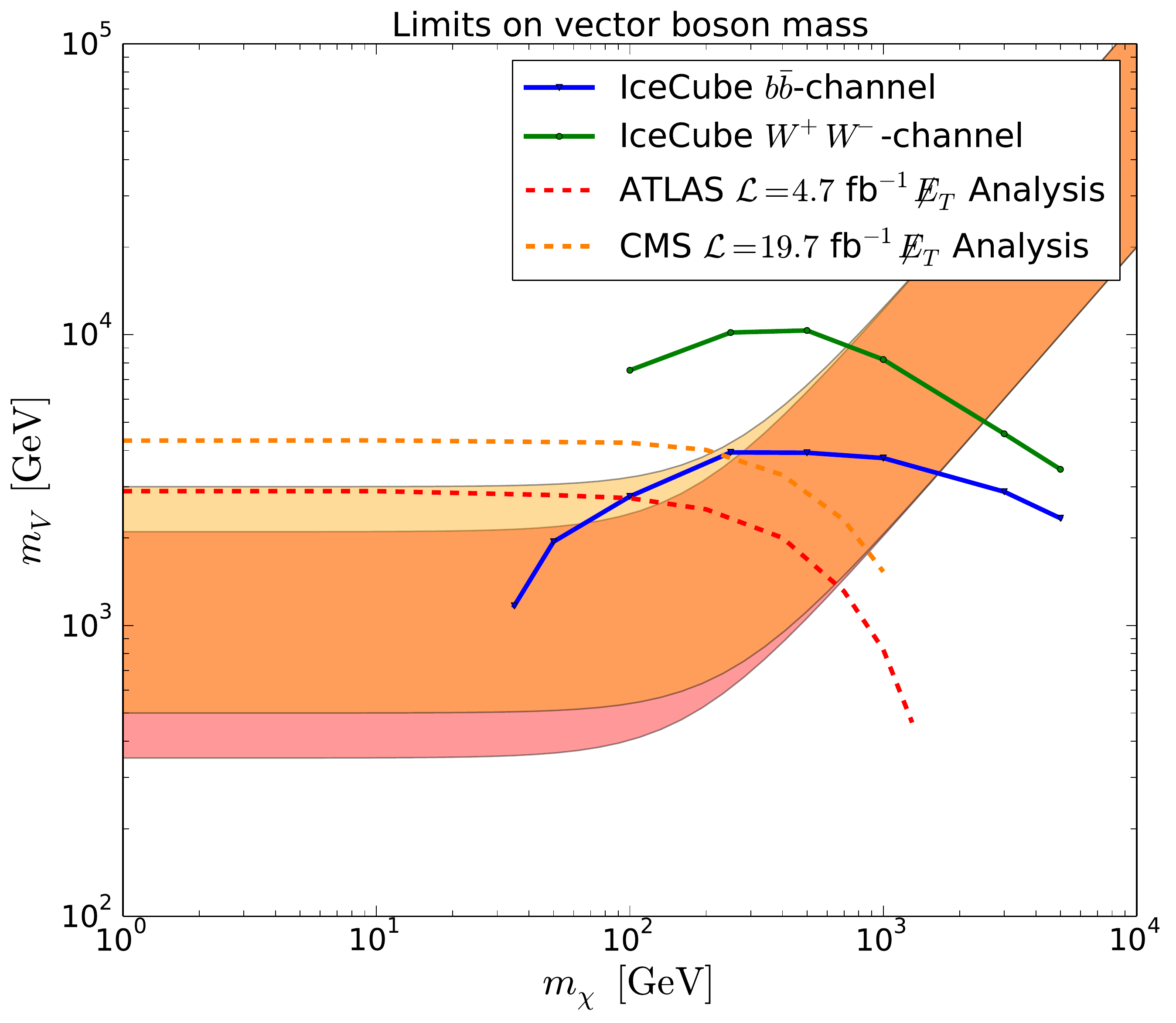}
 \caption[Upper limits on effective couplings and mediator mass in dark disc scenarios]{
The left (right) column shows the upper limits on the effective axial-vector coupling and vector mediator mass for Majorana dark matter derived from the IceCube solar WIMP analysis for dark matter disc scenarios with conservative (extreme) density fractions $\rho_{\rm disc}/\rho_{\rm halo} = 0.1 (1)$,  respectively. The blue (green) curve corresponds to limits set by the IceCube solar WIMP analysis for soft or hard neutrino spectra corresponding to the $b \bar b$-channel ($W^{+}W^{-}$-channel). The red (orange) dashed lines show upper limits from dark matter searches in mono-jet events at the LHC by the ATLAS (CMS) collaborations~\cite{Aad2013, CMS2014}. The gray band in the upper panels corresponds to the value of the coupling needed in order to account for the present day relic abundance of dark matter as measured by the Planck satellite \cite{Ade2013}. The black line indicates the upper limit for equilibrium between capture and annihilation in the Sun as derived from Eq.~(\ref{eq:equil}). The region above the dotted line corresponds to values of the parameters for which the effective field theory approach to dark matter annihilation is unreliable (see discussion in \ref{sec:results}). The red (orange) shaded area corresponds to the region given in \ref{eq:resonance}, where the effective field theory approach significantly underestimates the actual LHC limits provided by ATLAS (CMS), as described in Ref.~\cite{Buchmueller2013}.}
\label{fig:couplingmassdd}
\end{figure*}

Compared with the SMH, the dark matter disc velocity distributions contain a second peak at the value of the relative velocity of the disc. Even if the total local dark matter density is fixed to a value of $\rho_{\rm loc} = 0.3 \; \text{GeV} \, \text{cm}^{-3}$, the impact of this second WIMP population on the capture rate is significant. The resulting upper limits for a Majorana WIMP with universal coupling are shown in FIG.~\ref{fig:couplingmassdd} for the axial-vector operator $\mathcal{O}^{(6)}_{AA}$, Eq.~(\ref{eq:fermiwimp}). Note that the effect of the dark matter disc may be even stronger for lower relative velocities.

For the conservative estimate $\rho_{\rm disc}/\rho_{\rm halo} = 0.1$, the IceCube limits improve by a factor $1.5 - 2.4$, depending in detail on the mass of the WIMP. The region where IceCube can provide complementary limits to the searches at the LHC is enlarged down to WIMP masses of approximately $500 \; \text{GeV}$ for the $b \bar b$ benchmark channel, compared to approximately $850 \; \text{GeV}$ for the analysis without a co-rotating dark matter disk as presented in FIG.~\ref{fig:couplinglimits}. For the more extreme scenario $\rho_{\rm disc}/\rho_{\rm halo} = 1$, the limits improve by a factor $2.5 - 5$, and the point where the IceCube limits become more stringent than LHC limits is lowered further to approximately $250 \; \text{GeV}$. The same behavior is seen in the IceCube limits on mediator masses, lower panels in FIG.~\ref{fig:couplinglimits}, which increase to approximately $2.5 \; \text{TeV}$ ($4 \; \text{TeV}$) for the conservative (extreme) disc density fractions, assuming a soft neutrino spectrum corresponding to annihilation into $b\bar{b}$ final states.  It is important to note that the scenarios with a co-rotating dark matter disk would not lead to an increased signal in direct detection experiments, since these are mostly sensitive to the high-velocity part of the velocity distribution.


\section{Conclusion}\label{sec:conclusions}

We have presented a model-independent interpretation of the search for WIMP dark matter annihilation in the Sun with the IceCube Neutrino Observatory using an effective field theory formalism. 
If dark matter annihilation and capture in the Sun are in equilibrium, one can relate the IceCube upper limits on the annihilation rate to upper limits on the elastic WIMP-nucleon cross section. We have focused on spin-dependent interactions, to which the IceCube measurement is particularly sensitive, and have identified four higher-dimensional effective operators which give rise to spin-dependent non-relativistic WIMP-nucleon scattering: two operators for fermionic WIMPs, which reduce to one for Majorana fermions, and two operators for a vector boson WIMP. Scalar dark matter does not have spin-dependent interactions with quarks in the non-relativistic limit. 

We have constructed simplified models, where the interaction of dark matter and quarks is mediated by either a heavy scalar or a heavy vector particle. Predominantly spin-dependent interactions arise naturally in models with Majorana fermion dark matter and neutral vector mediators, leading to an effective axial-vector interaction, and in models with real vector dark matter and charged fermionic mediators with chiral couplings. 

We have calculated the spin-dependent elastic WIMP-proton scattering cross sections for these effective interactions and have placed upper limits on the coefficients of the effective operators from the IceCube data for scenarios with universal and Yukawa-like WIMP-quark couplings. The IceCube limits are sensitive to dark matter annihilation in a wide range of WIMP masses, from approximately 100\,GeV up to 5\,TeV. In the context of simplified models the upper limits on the coefficients of the higher-dimensional operators can be translated into lower limits on the mass of the mediating particle. For Majorana dark matter with universal axial-vector couplings to quarks, the IceCube measurement exclude vector mediators with masses below approximately 1\,TeV for WIMP masses up to 5\,TeV. 

We have discussed the astrophysical uncertainties of the IceCube limits on the effective operators from the local WIMP density and velocity distribution. Our results are not very sensitive to changes in the local dark matter density or deviations from the standard Maxwell-Boltzmann distribution of the dark matter velocity. However, the impact of a co-rotating dark matter disk, suggested by recent $N$-body simulations including the effect of baryons, could lead to a significant enhancement of the capture rate and increased sensitivity of the IceCue analysis.  

Finally, we have compared our limits to constraints from the dark mater relic density and direct searches at the LHC, and discussed the validity of the effective field theory approach in some detail. 
We have argued that the effective field theory interpretation of the IceCube data is theoretically reliable,  and also consistent with the dark matter relic density constraints. Furthermore, the IceCube analysis is complementary to Majorana dark matter searches at the LHC, extending the limits on higher-dimensional axial-vector operators to WIMP masses of  5\,TeV, \textit{i.e.}\ significantly beyond the current LHC reach. 


\begin{acknowledgments}
This work was supported by the Deutsche Forschungsgemeinschaft DFG through the graduate school ``Particle and Astroparticle Physics in the Light of the LHC'', by the Helmholtz Alliance for Astroparticle Physics (HAP), and by the U.S. Department of Energy under contract DE-AC02-76SF00515. MK would like to thank Thomas Rizzo and Matthew Dolan for discussions. 
\end{acknowledgments}


\bibliography{References}

\end{document}